\documentclass[apj]{emulateapj}

\usepackage{apjfonts}
\usepackage{lscape}

\usepackage{amsmath}
\usepackage{url}

\newcommand {\apgt} {\ {\raise-.5ex\hbox{$\buildrel>\over\sim$}}\ }
\newcommand {\aplt} {\ {\raise-.5ex\hbox{$\buildrel<\over\sim$}}\ }

\slugcomment{Accepted to ApJ}

\begin{document}

\title{Bayesian analysis to identify new star candidates in nearby young stellar kinematic groups}

\author{Lison Malo\footnote{Based on observations obtained at the Canada-France-Hawaii Telescope (CFHT) 
which is operated by the National Research Council of Canada, the Institut National des Sciences de 
l''Univers of the Centre National de la Recherche Scientique of France, and the University of Hawaii. }, 
Ren\'e Doyon, 
David Lafreni\`ere,
 \'Etienne Artigau,
 Jonathan Gagn\'e,
 Fr\'ed\'erique Baron}
\affil{D\'epartement de physique and Observatoire du Mont-M\'egantic, 
Universit\'e de Montr\'eal, Montr\'eal, QC H3C 3J7, Canada}
\and
\author{Adric Riedel}
\affil{Department of Astrophysics, American Museum of Natural History,
Central Park West at 79th Street, New York, NY 10024, USA}

%\altaffiltext{1}{Based on }
\email{malo@astro.umontreal.ca, doyon@astro.umontreal.ca, david@astro.umontreal.ca,  
artigau@astro.umontreal.ca, gagne@astro.umontreal.ca, baron@astro.umontreal.ca and
riedel@phy-astr.gsu.edu} 

\begin{abstract}
We present a new method based on a Bayesian analysis to identify new 
members of nearby young kinematic groups. 
The analysis minimally takes into account the position, proper motion, 
magnitude and color of a star, but other observables can be readily added 
(e.g. radial velocity, distance).
We use this method to find new
young low-mass stars in the $\beta$ Pictoris and AB Doradus 
moving groups and in the TW Hydrae, Tucana-Horologium, Columba, Carina 
and Argus associations.
Starting from a sample of 758 mid-K to mid-M (K5V-M5V) stars showing youth 
indicators such as H$\alpha$ and X-ray emission, our analysis yields 215 
new highly probable low-mass members
of the kinematic groups analyzed. One is in TW Hydrae, 37 in $\beta$ Pictoris, 
17 in Tucana-Horologium, 20 in Columba, 6 in Carina, 50 in Argus, 33 in AB Doradus, 
and the remaining 51 candidates are likely young but have an ambiguous membership to
more than one association. 
The false alarm rate for new candidates is estimated to be 
5\% for $\beta$ Pictoris and TW Hydrae, 10\% for Tucana-Horologium, 
Columba, Carina and Argus, and 14\% for AB Doradus. 
Our analysis confirms the membership of 58 stars proposed in the literature.
Firm membership confirmation of our new candidates will require 
measurement of their radial velocity (predicted by our 
analysis), parallax and lithium 6708 \AA\ equivalent width. 
We have initiated these follow-up observations for a number of 
candidates and we have identified 
two stars (\objectname{2MASSJ01112542+1526214}, \objectname{2MASSJ05241914-1601153})
as very strong candidate members of the $\beta$ Pictoris moving group and one strong candidate 
member (\objectname{2MASSJ05332558-5117131}) of the
Tucana-Horologium association; these three stars have radial velocity 
measurements confirming their
membership and lithium detections consistent with young age. 
Finally, we proposed that six stars should be considered as 
new {\em bona fide} members of $\beta$PMG and ABDMG, one of which being first 
identified in this work, the others being known candidates from the literature.
\end{abstract}


%%%%%%%%%%%%%%%%%%%%%%%%%%%%%%%%%%%%%%%%%%%%%%%%%%%%%%%%%%%%%%%%%%%%%%%%%%%%%%%%%%%%%%%%%%%%%%%%%%%%%%%%%%%%%%%%%%%%%%%
\section{INTRODUCTION}
%%%%%%%%%%%%%%%%%%%%%%%%%%%%%%%%%%%%%%%%%%%%%%%%%%%%%%%%%%%%%%%%%%%%%%%%%%%%%%%%%%%%%%%%%%%%%%%%%%%%%%%%%%%%%%%%%%%%%%%
Nearby young co-moving groups are sparse, gravitationally unbound stellar 
associations comprising a few dozens of stars scattered within $\sim$ 
100\,pc of the Sun with ages ranging from 5 to a few hundred Myr. 
Co-moving group members are characterized by a common position and space
motion within the Galaxy. 
As a result of a projection effect, they display an organized motion 
on the sky moving towards a convergent point \citep[apex;][]{2001montes} 
and this can be used to discriminate genuine members from field stars. 
To identify new associations and/or new members of a given association, 
\citet{1958eggen,1995eggen} used the convergent point criterion, which is based 
on the direction of motion of group members on the celestial sphere, together with 
new criteria based on the amplitude of the stars' motion. 
The latter are known as Eggen's criteria.
Eggen's method was slightly modified by \citet{2001montes} to take into 
account the 3D space motion dispersion of the stars. 
They used proper motion, radial velocity and distance uncertainties 
to measure the galactic space velocities and their dispersion among 
the members of an association.
Generally, proper motion, radial velocity and distances constitute 
minimal information needed for identifying young stars from co-moving 
groups.

In the last decade, several new associations have been identified, thanks
mostly to Hipparcos \citep{1997perryman} and other 
large scale surveys like TYCHO \citep{2000hog} and 2MASS \citep{2006skrutskie}. 
\citet{2001bzuckerman} improved the method to confirm candidate membership to 
young associations by 
including constraints on their photometric properties.
Following the advent of the 2MASS PSC catalog \citep{2003cutri}, \citet{2003song} 
used an optical-infrared color-magnitude diagram to better discriminate 
between K and M dwarfs. 
The Hipparcos catalog made it possible to determine space velocities 
with good precision for approximately 20\% of nearby young star candidates. 
For the remaining 80\%, \citet{2004song} used the ``good box'' method 
for finding other candidate young stars with photometric 
and kinematics properties similar to the known members.
Similarly, \citet{2006torres,2008torres} led the development of a merit function 
for identifying new members (Search for Associations Containing Young stars; SACY).

Although significant progress has been made in the last decade in finding 
young stars in co-moving groups, their identification remains 
challenging because they are sparsely dispersed over the celestial 
sphere. 
Moreover, the studies mentioned above have uncovered relatively few 
low-mass (K and M) stars in nearby young associations since they relied 
mostly on surveys in the $V$ band with a limiting magnitude around $V$=14.
However, some recent studies are now providing a substantial number of 
candidates in the low-mass regime \citep{2006torres,2008torres,2009lepine,2010schlieder,2011shkolnik,2011desidera,2011kiss,2011rodriguez,2011riedel,2012schlieder,2012shkolnik}. 

In this paper, we present a new method for identifying young low-mass
stars in kinematic groups. 
More specifically, the method allows calculating an objective 
membership probability of a given candidate to an ensemble of moving groups. 
This method is applied to the seven youngest kinematic groups 
closest to the Sun. 
The paper is structured as follows. 
A brief description of the young associations analyzed 
is presented in $\S$2 and more details on the kinematic properties, space 
location and photometric properties of the known members of these 
associations are presented in $\S$3. 
A presentation of the kinematic model describing these associations
follows in $\S$4. 
The Bayesian method used for selecting new members is described in $\S$5.
Then the sample to which our method is applied is described in $\S$6, followed by
the results is $\S$7. Follow-up observations (radial velocity and lithium absorption)
 for some candidates are presented and discussed in $\S$8. 
Section\,$\S$9 is devoted to a discussion of previously known members and new 
candidate members unveiled from this work. 
Concluding remarks and suggestions for future work follow in $\S$10.

%%%%%%%%%%%%%%%%%%%%%%%%%%%%%%%%%%%%%%%%%%%%%%%%%%%%%%%%%%%%%%%%%%%%%%%%%%%%%%%%%%%%%%%%%%%%%%%%%%%%%%%%%%%%%%%%%%%%%%%
\section{THE YOUNG ASSOCIATIONS} \label{chap:deux}
%%%%%%%%%%%%%%%%%%%%%%%%%%%%%%%%%%%%%%%%%%%%%%%%%%%%%%%%%%%%%%%%%%%%%%%%%%%%%%%%%%%%%%%%%%%%%%%%%%%%%%%%%%%%%%%%%%%%%%%
Our search for young low-mass stars will be restricted to the seven closest
co-moving groups: the $\beta$ Pictoris Moving Group ($\beta$PMG), 
the TW Hydrae Association (TWA), the Tucana-Horologium Association (THA), 
the Columba Association (COL), the Carina Association (CAR), 
the Argus Association (ARG) and the AB Doradus Moving Group (ABDMG). 
The basic properties of these associations, which are detailed below, 
are summarized in Tables~\ref{tab:prop1} and ~\ref{tab:prop2}.
Thereafter in this paper, we consider as {\em bona fide members} of 
young kinematic groups
all stars with a good measurement of trigonometric distance, proper motion,  
Galactic space velocity and other youth indicators 
such as H$\alpha$ emission, X-ray emission, appropriate location in the 
HR diagram and lithium absorption; those 177 stars are listed in Table~\ref{tab:membprop}. 

\begin{deluxetable}{lccc}
\tablewidth{0pt}
\tablecolumns{4}
\tablecaption{Properties of young Local associations \label{tab:prop1} }
%\tabletypesize{\Large}
\tablehead{
\colhead{Name} & \colhead{Age} & \colhead{Distance} & \colhead{Number} \\
\colhead{of group} & \colhead{range (Myr)} & \colhead{range (pc)} & \colhead{of stars\tablenotemark{a}}}
\startdata
$\beta$ Pictoris ($\beta$PMG) & 12-22 & 9-73 & 39\\
TW Hydrae (TWA) & 8-20 & 28-92 & 10\\
Tucana-Horologium (THA) & 10-40 & 36-71 & 44\\
Columba (COL) & 10-40 & 35-81 & 21\\
Carina (CAR) & 10-40 & 46-88 & 5\\
Argus (ARG) & 30-50 & 8-68 & 11\\
AB Doradus (ABDMG) & 50-120 & 7-77 & 47
\enddata
\tablenotetext{a}{Members with published trigonometric distance only.}
\end{deluxetable}

\subsection{$\beta$ Pictoris moving group}

This group was proposed by \citet{2001bzuckerman} following the work of 
\citet{1999barrado}.
Members sharing the galactic motion of its namesake star were selected from 
the Hipparcos \citep{1997perryman} and 
\citet{1991gliese} databases.
The age of this association is estimated from color-magnitude diagrams \citep[$20\pm 
10$\,Myr;][]{1999barrado}, stellar formation models \citep[$22\pm12$\,Myr;][]{2007makarov} 
and Li abundance \citep[$12\pm8$\,Myr;][]{2001bzuckerman}.
\citet{2008fernandez} summarize the estimated ages published in the 
literature (see their Table 2). 
At present, the $\beta$PMG counts 39 {\em bona fide members}
\citep{2004zuckerman,2009teixeira,2010rice,2010schlieder,2011kiss,2012faherty,2012schlieder} 
ranging from 9 to 73\,pc. 
\citet{2010rice} have identified the lowest mass, isolated, member of $\beta$PMG: 
an M8.5V brown dwarf. \citet{2012faherty} have measured the trigonometric distance
of this object and this mesurement confirms its membership to this group.
The members are scattered over the celestial sphere with a majority of 
them in the Southern hemisphere.
The observational properties of all {\em bona fide members} are listed in
Table~\ref{tab:membprop}.

\citet{2006torres} and \citet{2008torres} identified respectively 9 and 6 new 
candidate members of this association from the minimization of a merit function 
described by space velocities, galactic positions
and theoretical isochrones ($M_{V}$ vs $V-I$). 
\citet{2009lepine} and \citet{2010schlieder} each introduced 4 
new candidate members, selected from the LSPM \citep{2005lepine} and
TYCHO-2 catalogs according to kinematic and optical/IR photometric criteria.
\citet{2011kiss} proposed 5 new candidate members using the RAVE 
\citep{2006steinmetz,2008zwitter} and Hipparcos catalogs. 
\citet{2010nakajima} and \citet{2012nakajima} revisited 
the Hipparcos catalog with a new method to identify members to stellar kinematic
groups within 20\,pc and 30\,pc of the Sun, respectively; they proposed 5 
candidates of $\beta$PMG within 30\,pc.
Recently, \citet{2012schlieder} identified 2 likely new members with a 
constistent Hipparcos distance.

\subsection{TW Hydrae association}

TWA was the first nearby association discovered, 
in the studies of \citet{1989delareza} and \citet{1992gregorio} following the 
work of \citet{1983rucinski} on TW Hya.
These studies used the IRAS \citep{1988helou} point-source catalog to 
identify four other T Tauri systems in the same region of the sky \citep{1999webb}.
\citet{1997kastner} concluded that TWA forms a physical association with five 
strong X-ray emitters. 
\citet{1999webb} associated six more stellar systems (7 stars and 1 brown dwarf) 
to TWA using the X-ray properties of all stars within $12\,^{\circ}$ around TW Hya.
TWA now counts around 30 potential members \citep{2005mamajek} including 10 
{\em bona fide members} (see Table~\ref{tab:membprop}). 
Recently, \citet{2012nakajima} proposed 7 candidates within 30\,pc.

The age of TWA is estimated from HR diagram along with pre-main sequence 
tracks \citep[8\,Myr;][]{1999webb}, 
HR diagram with H$\alpha$ and lithium equivalent width 
\citep[$10^{+10}_{-7}$\,Myr;][]{2006barrado}
and the expansion age \citep[$20^{+25}_{-7}$\,Myr;][]{2005mamajek}. %voir Fernandez2008
Using galactic dynamics, \citet{2006delareza} determined an age of 8$\pm$0.8\,Myr.
\citet{2000mamajek} and recently \citet{2012song} suggested that TWA is likely the near edge of a larger 
population (i.e., Lower Centaurus Crux).

\subsection{The Great Austral Young Association (GAYA)}
\citet{2003torres,2006torres} identified many young stars with similar 
kinematic and photometric properties.
This group of stars was proposed to be called the Great Austral Young Association 
(GAYA) and was later sub-divided into three groups: THA, COL and CAR \citep{2008torres}.

\subsubsection{Tucana-Horologium association}

Discovered simultaneously, the Tucana association 
\citep*{2000zuckermanwebb} and the Horologium association 
\citep{2000torres} were combined together because of their similar
space motion, age and distance \citep{2001reza}.
In the first case, \citet*{2000zuckermanwebb} used 
the Hipparcos and the IRAS catalogs to find 10 stars having a 
similar galactic motion in both catalogs. 
The Horologium association was identified by \citet{2000torres} 
using the ROSAT \citep{1994voges} catalog in order to find X-ray 
emitting sources around the active star EP Eri.

The following work of \citet{2003song} discovered 11 new members 
with a galactic motion similar to those discovered in previous studies.
THA counts 44 {\em bona fide members} \citep{2004zuckerman,2008torres,2011kiss,2011zuckerman} 
ranging from 36 to 71\,pc (see Table~\ref{tab:membprop}). 
More recently, \citet{2008torres} and \citet{2011kiss} 
respectively identified 9 and 2 new candidate members.
\citet{2011zuckerman} proposed 8 new members to THA with spectral types between 
A1V and G8V.
\citet{2012nakajima} proposed 4 more candidates within 30\,pc.

The age of THA is estimated at 40\,Myr \citep*{2000zuckermanwebb}
based on various age indicators including Li equivalent width, H$\alpha$ profile 
and gyrochronology.
The kinematic evolution of the association with time provides further constraints 
on its age.
Assuming a coeval formation with an initial velocity dispersion of 
1.5\,km s$^{-1}$, the group of stars should have dispersed over 70\,pc 
after a period of 20\,Myr \citep{2001torres}. 
From the studies of \citet{2000stelzer} and \citet{2001tzuckerman}, they estimate 
the lower age limit at 10\,Myr. 

\subsubsection{Columba association}

COL includes 53 proposed members by \citet{2008torres,2011zuckerman}, 
with 21 listed as {\em bona fide members} (see Table~\ref{tab:membprop}). 
Three stars proposed by \citet{2008torres} to be members
of COL were already proposed by \citet{2004zuckerman} to be members of THA. 
%We will discuss these stars later (see~section~\ref{chap:neuf}).
Recently, \citet{2011zuckerman} added 14 new members (B9V to M0.5V) including 
multiple planet host star HR 8799 \citep{2008marois} whose membership was also 
confirmed by \citet{2010doyon} through a Bayesian analysis similar to the 
one presented in this paper.
The age of COL is estimated to be similar to that of THA \citep{2008torres}.

\subsubsection{Carina association}
This association was discovered by the same method as THA and COL \citep{2008torres}.
\citet{2008torres} proposed 23 members, with only five listed as {\em bona fide members}.
(see Table~\ref{tab:membprop}). They also showed that the stars in CAR have 
properties and age similar to THA and COL members.
Given the small number of {\em bona fide members} identified in CAR, this association 
cannot be considered as well defined as others like $\beta$PMG, THA and COL.

\subsection{Argus association}
This association was initialy unveiled by \citet{2000makarov} using proper motion
to identify a bulk of stars located in the Carina-Vela moving group.
The IC 2391 open cluster was proposed to be part of this moving group by 
\citet{2000makarov}, and
\citet{2003torres} used the convergence point method to show that the kinematics of 
these two groups are similar. %\citet{2008torres}
\citet{2003torres} identified this group as the Argus Association due to its 
special galactic space velocity $U$ and proposed a different definition for this group. 
\citet{2011riedel} proposed that the closest (8.4\,pc) pre-main sequence 
star should be AP Col, a member of the Argus association.
\citet{2011zuckerman} proposed 6 new members with a spectral type between A0V and F4V.
Recently, \citet{2011desidera} proposed HIP 36948 as a candidate member of Argus
association from several age-dating indicators.
ARG counts 11 {\em bona fide members} (see Table~\ref{tab:membprop}).
This association is not as well defined as other groups because the properties
of all known members are not well measured. 
There is no parallax mesurement for the IC 2391 members.

The age of the ARG members is similar to that of the IC 2391 members 
as inferred from the lithium equivalent width and position in the HR diagram \citep[40\,Myr;][]{2008torres}. 
An age of 50$\pm$5\,Myr was inferred by \citet{2004barrado} using lithium 
depletion and H$\alpha$ emission.
 
\subsection{AB Doradus moving group}

This group was identified by \citet{2004zuckerman} by performing a kinematic 
analysis of the Hipparcos catalog.
Forty-seven {\em bona fide members} are associated with this group \citep{2004zuckerman,2008torres,2011zuckerman}
and range in distance from 7 to 77\,pc. 
Several new candidate members have been proposed recently: 43 by \citet{2008torres}, 
6 by \citet{2010schlieder}, 6 by \citet{2012schlieder} and 1 by \citet{2012bowler}. 
\citet{2012nakajima} proposed 8 (2 F and 6 M stars) more candidates of 
ABDMG within 30\,pc.

In addition to having a common motion, the members show signs of 
youth, such as strong H$\alpha$ emission and/or the presence of Li.
The observational properties of all members are listed in 
Table~\ref{tab:membprop}.

The age of ABDMG is estimated by various age indicators.
An age of $\sim$ 50\,Myr is deduced  
by a color-magnitude diagram (CMD)
($M_{V}$ vs $V-K$; see figure 1 of \citealt{2004bzuckerman}).
\citet{2005luhman} compared the same diagram for ABDMG and the open cluster 
IC 2391 (30-50\,Myr) to deduce an age between 75 and 150\,Myr. 
Using Li equivalent widths and the CMD $M_{V}$ vs $V-I$, \citet{2006lopez} 
estimated different ages for
two subgroups of ABDMG: 30-50\,Myr and 80-120\,Myr. 
Recently, ages of 45\,Myr (lower limit) and 70\,Myr were
estimated from Li equivalent widths by \citet{2008mentuch} and \citet{2009dasilva}, respectively. 

%%%%%%%%%%%%%%%%%%%%%%%%%%%%%%%%%%%%%%%%%%%%%%%%%%%%%%%%%%%%%%%%%%%%%%%%%%%%%%%%%%%%%%%%%%%%%%%%%%%%%%%%%%%%%%%%%%%%%%%
\section{GLOBAL PROPERTIES OF KNOWN MEMBERS AND FIELD STARS} \label{chap:trois}
%%%%%%%%%%%%%%%%%%%%%%%%%%%%%%%%%%%%%%%%%%%%%%%%%%%%%%%%%%%%%%%%%%%%%%%%%%%%%%%%%%%%%%%%%%%%%%%%%%%%%%%%%%%%%%%%%%%%%%%

\begin{deluxetable*}{lrrrr}
\tabletypesize{\scriptsize}
\tablewidth{0pt}
\tablecolumns{5}
\tablecaption{Mean galactic motion and position \label{tab:prop2}}
\tablehead{
\colhead{Name} & \colhead{$UVW$} & \colhead{$\sigma_{\sc UVW}$} & \colhead{$XYZ$} & \colhead{$\sigma_{\sc XYZ}$} \\
\colhead{of group} & \colhead{(km s$^{-1}$)} & \colhead{(km s$^{-1}$)} & \colhead{(pc)} & \colhead{(pc)}
}
\startdata
$\beta$ Pictoris ($\beta$PMG) & $-10.94, -16.25, -9.27$ & $2.06, 1.30, 1.54$ & $9.27, -5.96, -13.59$ & $31.71, 15.19, 8.22$ \\
Tucana-Horologium (THA) & $-9.88, -20.70, -0.90$ & $1.51,1.87,1.31$ & $11.39, -21.21,-35.40$ & $19.29,9.17,5.39$ \\
AB Doradus (ABDMG) & $-7.12,-27.31,-13.81$ & $1.39,1.31,2.16$ & $-2.37,1.48,-15.62$ & $20.03,18.83,16.59$ \\
Columba (COL) & $-12.24,-21.32,-5.58$ & $1.03,1.18,0.89$ & $-27.44,-31.32,-27.97$ & $13.79,20.55,15.09$ \\
Carina (CAR) & $-10.50,	-22.36,-5.84$ & $0.99,0.55,0.14$ & $15.55,-58.53,-22.95$ & $5.66,16.69,2.74$ \\
TW Hydrae (TWA) & $-9.87,-18.06,-4.52$ & $4.15,1.44,2.80$ & $12.49,-42.28,21.55$ & $7.08,7.33,4.20$ \\
Argus (ARG) & $-21.78,-12.08,-4.52$ & $1.32,1.97,0.50$ & $14.60,-24.67,-6.72$ & $18.60,19.06,11.43$ \\
Field stars & $-10.92,-13.35,-6.79$ & $23.22,13.44,8.97$ & $-0.18,2.10,3.27$ & $53.29,51.29,50.70$
\enddata
\end{deluxetable*}

This section summarizes the main kinematic and photometric 
properties for the known members of the seven associations studied here. 
As the Bayesian analysis performed later requires considering also that a star 
belongs to the field, as an hypothesis, 
the properties of field stars are also discussed.
These properties will be used in the next section for 
constructing kinematic and photometric models
for each association.
For simplicity, we model these properties using simple functions: 
the Galactic space velocities and positions distributions are modeled by 
Gaussians whose parameters are inferred
by fitting cumulative distribution functions while the 
color magnitude sequences are modeled with polynomials.

\subsection{Kinematic properties}
\begin{figure}[!hbt]
\epsscale{1.1}
\plotone{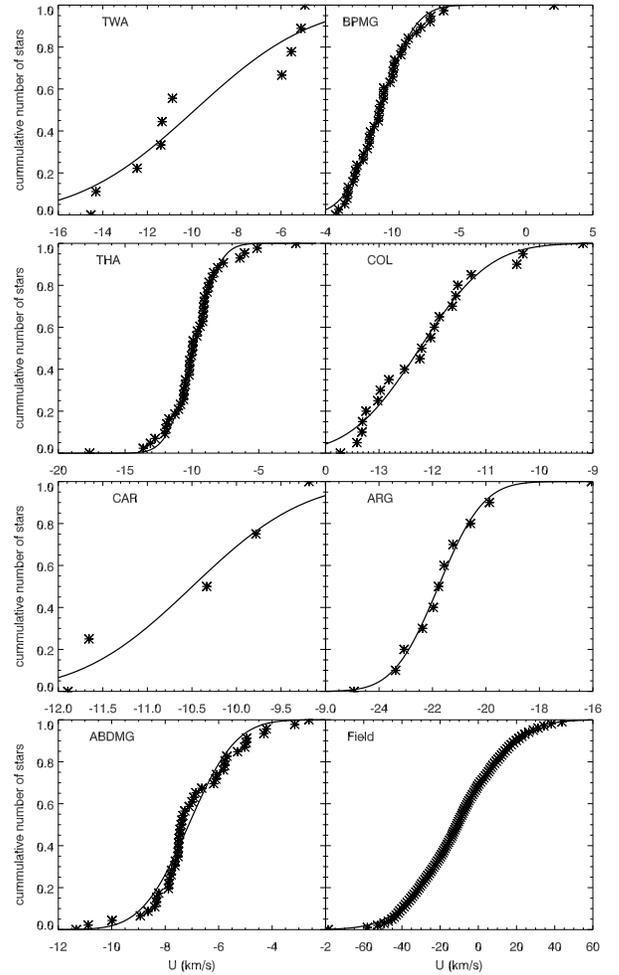}
\caption{\footnotesize{$U$ cumulative distribution functions of {\em bona fide members} 
of young kinematic groups and the field stars. 
The black line represents the adopted parametrization 
(see Table~\ref{tab:prop2}).} \label{fig:fig1}}
\end{figure}

Because of the coeval formation of a given association, all members 
share, to within a few km s$^{-1}$, a common space velocity within 
the Galaxy.
The galactic space motion of a star, $UVW$, is determined from 
its sky position ($\alpha$, $\delta$), radial velocity, proper 
motion and parallax, using the \citet*{1987johnson} relations.
Proper motion and parallax measurements come from several sources 
\citep[][; Riedel in prep.]{2007vanleeuwen,2007gizis,2008teixeira,2009teixeira,2011riedel,2012faherty}, while radial 
velocities are taken from various studies 
(see Table~\ref{tab:membprop}).
The field star sample we used comprises 10094 stars within 150 pc 
with Hipparcos parallax known to an accuracy better than 
5$\sigma$ (from \citealp{2007vanleeuwen}) and 
radial velocities from \citet{2009francis}.
Figure~\ref{fig:fig1} shows the $U$ cumulative distribution function 
for the seven associations and for the field stars sample, as an example. 
These distributions are reasonably well approximated by normal functions whose 
parameters are given in Table~\ref{tab:prop2}.

\subsection{Galactic position}

\begin{figure}[!hbt]
\epsscale{1.1}
\plotone{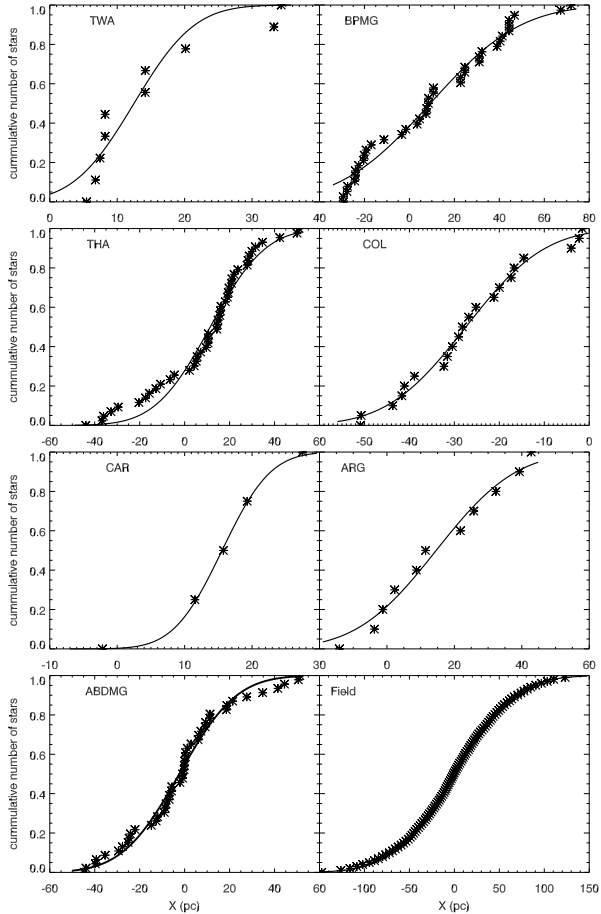}
\caption{\footnotesize{$X$ cumulative distribution functions of the known members
of young kinematic groups and the field stars. 
The black line represents the adopted parametrization 
(see Table~\ref{tab:prop2}).} \label{fig:fig2}}
\end{figure}

By virtue of their youth and their coeval formation, the members of young
associations have had little time to disperse within the Galaxy. 
As a result, the positions $XYZ$ of the members of a young association are 
relatively well confined within the Galaxy.
The $XYZ$ frame of reference is centered on the position of the Sun and
follows the same sign convention as the galactic velocities $UVW$, i.e. 
$X$ positive towards the center 
of the Galaxy, $Y$ positive in the direction of galactic rotation 
and $Z$ positive towards the north galactic pole.
Table~\ref{tab:prop2} gives the mean values and dispersions for
the $XYZ$ distribution of the young kinematic group members and field stars.
The $X$ cumulative distributions of the known members of the seven associations
and the field stars are presented in figure~\ref{fig:fig2}.

\subsection{Photometric properties}

CMDs have been a crucial tool for identifying 
young stars. Since our goal is to search for low-mass stars, color
indices as red as possible are desirable and we opted for the $I_{c}-J$ index.
A color index based on $V$-band magnitudes, largely used in previous 
studies to search
for young stars, is impractical for many objects later than M5V.

For the {\em bona fide members} and the field stars, the $I_{c}$ magnitudes come
from Hipparcos \citep{2011anderson}, DENIS \citep{1997epchtein}, SDSS-DR8 \citep{2011adelman} and other studies.
We transformed the Gunn-$i$ DENIS and SDSS magnitudes to $I_{c}$ 
using conversions derived using the standards of Landolt \citep{2009landolt}. These transformations are 
\begin{alignat}{2}
I_{c} &= i_{DENIS} + 0.01 \\
I_{c} &= i_{SDSS-DR8} - 0.67.
\end{alignat}
and accurate within 0.2 mag over 
-0.10 < $I_{c}-J$ < 3.25.

The $J$ magnitude is taken from the 2MASS PSC \citep{2003cutri}. 
Field stars are taken from the samples of \citet*{2009francis} 
and \citet{2003phanbao}.
The first includes 10094 stars within 150 pc with radial velocity and 
trigonometric distance measurements. 
The second includes 64 M dwarfs with parallax measurements from various studies.

The least massive known members of the seven young groups studied are 
M8.5V ($\beta$PMG and TWA) and M3V (ABDMG) dwarfs. 
Thus, to extend the color sequence of the young associations beyond M3V 
up to M9V, we used the evolutionary models 
of \citet{1998baraffe,2002baraffe}, shifting the 
isochrones (typically by 0.5 mag) to match the 
known members between K5V and M5V dwarfs. 

For the purpose of the current analysis, the association members are merged into
four groups: $<$10\,Myr-old (TWA), 10\,Myr to 20\,Myr-old ($\beta$PMG), 
20\,Myr to 50\,Myr-old (THA, COL, CAR and ARG) and $>$50\,Myr-old (ABDMG).
Since ABDMG lacks massive (A0V dwarf) known members, a sample of 
33 early-type Pleiades members (with Hipparcos parallax) were used to reproduce 
the trend of the color sequence from A0V to M9V for the ABDMG.

\begin{figure}[htb!]
\epsscale{1.2}
\plotone{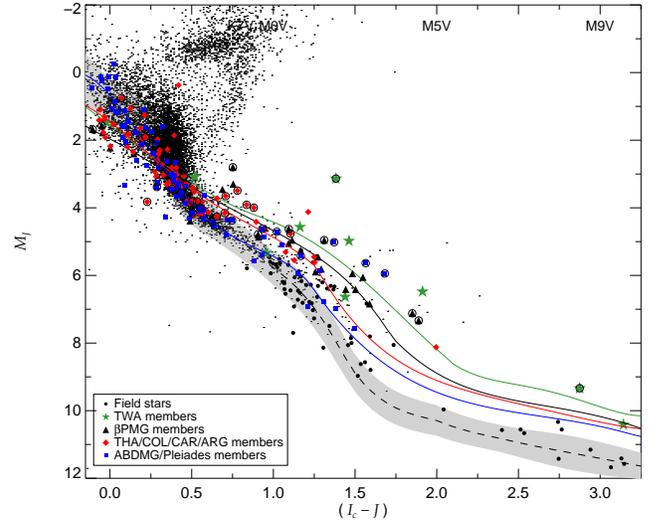}
\caption{\scriptsize{Color-magnitude diagram ($M_{J}$ vs $I_{c}-J$) for members 
of TWA (green stars), $\beta$PMG (black triangles), THA, COL, CAR and ARG
(red diamonds) and ABDMG and Pleiades (blue squares). Fields stars (dots and filled black 
circles) are from \citet*{2009francis} 
and \citet{2003phanbao}. The dashed line and shaded area represent the locus 
of old field stars. K5V-M5V, representative of our search sample (see $\S$6), 
have $0.8 < I_{c}-J < 2.0$. On average, young
late-type stars are brighter than field stars, a property that can be 
used, along with other kinematic properties, to discriminate young stars 
from old ones. Binary stars are those with black circles superposed on their own symbol.} \label{fig:jabsconnu}}
\end{figure}

We used the 8\,Myr model for TWA, 12\,Myr for $\beta$PMG and 
40\,Myr for the THA, COL, 
CAR and ARG groups.
The oldest groups (including ABDMG and Pleiades members) were approximated 
as a 80\,Myr model. 
For the field star sample, we used the 5\,Gyr model. 
We corrected the magnitudes of the models to match the 2MASS color system.
All known binaries were excluded for the determination of the color sequences.
Figure~\ref{fig:jabsconnu} shows the resulting CMD sequences adopted, for the 
association members and old field stars. 
For $I_{c}-J > 0.8$, young stars are significantly overluminous compared to 
field stars. 
The absolute magnitude $M_{J}$ of those four groups are well described, 
within a dispersion of 0.3\,mag, using polynomials.
Similarly, one can construct an empirical sequence for old field stars. 
This sequence is represented by the dashed line in 
Figure~\ref{fig:jabsconnu}. 
The 1$\sigma$ dispersion, represented by the grey envelope, varies with color and 
is typically $\sim$0.5\,mag.

%%%%%%%%%%%%%%%%%%%%%%%%%%%%%%%%%%%%%%%%%%%%%%%%%%%%%%%%%%%%%%%%%%%%%%%%%%%%%%%%%%%%%%%%%%%%%%%%%%%%%%%%%%%%%%%%%%%%%%%
\section{KINEMATIC MODEL} \label{chap:quatre}
%%%%%%%%%%%%%%%%%%%%%%%%%%%%%%%%%%%%%%%%%%%%%%%%%%%%%%%%%%%%%%%%%%%%%%%%%%%%%%%%%%%%%%%%%%%%%%%%%%%%%%%%%%%%%%%%%%%%%%%

A key element of our analysis for identifing new members of 
young associations is to build a kinematic model of a given association.
For a star at a given position on the sky and given the mean and dispersion of
the galactic space velocity of an association, 
this model should reliably predict the radial and tangential 
velocities and the direction of proper motion that the star would have
if it were an actual member of the association.
For a given distance, the tangential velocity translates into a proper motion amplitude.
The direction and amplitude of the 
proper motion predicted by this model restrict considerably 
the number of potential members of a young association, 
and even help constraining their distance.

The kinematic model is built by inverting the procedure described in 
Section~\ref{chap:trois}.
For a specific group and a position in the sky (right ascension and 
declination), we create virtual stars having $UVW$ velocities and dispersions 
as given by Table~\ref{tab:prop2}. The radial and 
tangential velocities of these stars are calculated and their mean and
dispersion are obtained.
Then given a distance, the tangential velocity is used to estimate 
the amplitude of the proper motion.

\begin{figure}[!hbt]
\epsscale{1.2}
\plotone{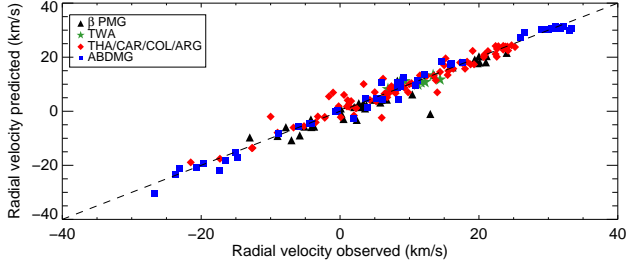}
\caption{\footnotesize{Comparison between estimated radial velocities by the 
kinematic model and those observed for the known members of 
$\beta$PMG (black triangles), TWA (green stars), 
THA,COL,CAR,ARG (red asterisks) and ABDMG (blue diamonds).}\label{fig:vradconnu}}
\end{figure}

The robustness of this kinematic model can be gauged by comparing 
the measured and predicted radial velocities for {\em bona fide members}. 
This comparison is presented in figure~\ref{fig:vradconnu} for 
the seven associations.
There is an excellent correlation between the predicted and 
observed values, with an rms of 1.9 km s$^{-1}$.

%%%%%%%%%%%%%%%%%%%%%%%%%%%%%%%%%%%%%%%%%%%%%%%%%%%%%%%%%%%%%%%%%%%%%%%%%%%%%%%%%%%%%%%%%%%%%%%%%%%%%%%%%%%%%%%%%%%%%%%
\section{SELECTION OF CANDIDATES - BAYESIAN STATISTICAL ANALYSIS}  \label{chap:cinq}
%%%%%%%%%%%%%%%%%%%%%%%%%%%%%%%%%%%%%%%%%%%%%%%%%%%%%%%%%%%%%%%%%%%%%%%%%%%%%%%%%%%%%%%%%%%%%%%%%%%%%%%%%%%%%%%%%%%%%%%

\subsection{General formalism} 
Based on a set \{${\theta}$\} of observables, a Bayesian analysis is 
used to select stars potentially members of a young kinematic group. 
From this analysis, a membership probability and a statistical distance
are determined for each star and each group considered.
In this study, the observables are the amplitude of proper motion in 
right ascension and declination, the apparent $I_{c}$ and $J$ magnitudes and the position 
($\alpha$, $\delta$) of the star on the sky.
It is possible to add more observables in the analysis 
(e.g, radial velocity), but in practice they are limited.
Let \{$H_k$\} be a set of hypotheses $H$$^{g_{m}}_{d_{n}}$ 
\begin{equation} \label{ensh}
\{H_{k}\}=\{ (H^{g_{1}}_{d_{1}}), (H^{g_{1}}_{d_{2}}), ..., (H^{g_{1}}_{d_{n}}), (H^{g_{2}}_{d_{1}}), (H^{g_{2}}_{d_{2}}), ..., (H^{g_{m}}_{d_{n}}) \}
\end{equation}
that a candidate is a member 
of a group ($g_{m}$) at a given distance $d_{n}$.
The set contains $m \times n$ hypotheses where $m$ represents the number of groups 
considered and $n$ the number of distances. 
Given the observable $\theta$, the probability that 
a candidate star is a member of group $g_{i}$ at distance $d_{j}$ is $P(H^{g_{i}}_{d_{j}}|$$\theta)$.
In what follows, the symbol ``|'' means ``given''.

According to Bayes' theorem, we have:
\begin{eqnarray}
\label{baye1}
P( H^{g_{i}}_{d_{j}} | \theta) &=& \frac {P (\theta | H^{g_{i}}_{d_{j}} ) P(H^{g_{i}}_{d_{j}}) }{P (\theta)}
\end{eqnarray}
with, 
\begin{eqnarray}
P(\theta) &=& \sum_{k=1}^{m \times n}P(\theta|H_{k})P(H_{k})  \nonumber
\end{eqnarray}
where $P(\theta|H_{k})$ is the probability to obtain the observable 
$\theta$ under a given hypothesis.
This quantity can be easily determined with a 
model representative of the observable $\theta$. 
In the denominator of Equation~(\ref{baye1}), 
$P(\theta)$ is the marginal probability to obtain 
the observable $\theta$ independently of the
hypothesis considered. This term normalizes the numerator by 
summing over all possibilities. $P(H_k)$ is the prior probability that the hypothesis is true.
Since these prior probabilities are unknown, an equal weight 
is given to them: $P(H_k)$ = ($1/m \times n$).  

Eq.~(\ref{baye1}) is a probability density function whose
maximum gives the most likely hypothesis and the most likely 
distance (hereafter referred to as the {\sl statistical} distance), while
its sum over all distances (for each group hypothesis)
gives the membership probability irrespective of the distance :
\begin{equation} \label{totbaye}
P(H^{g_{i}}|\theta)=\sum_{j=1}^{n}P( H^{g_{i}}_{d_{j}} | \theta).
\end{equation}

As mentioned before, the probabilities $P(\theta | H_{k})$ needed 
in Eq.~(\ref{baye1}) are derived from a 
Gaussian distribution with mean $\bar{\theta}$ and standard deviation 
$\sigma$:

\begin{equation} \label{pabaye}
P (\theta | H) = \frac {1}{\sqrt{2\pi} \sigma} e^{-\frac{1}{2} \left(\frac{(\theta - \overline{\theta})}{\sigma}\right)^2}
\end{equation}

By considering two observables and applying Baye's theorem iteratively, equation~(\ref{baye1}) becomes:

\begin{equation} 
\label{bayetotal}
P(H^{g_{i}}_{d_{j}} | \theta_{1} \cap \theta_{2}) = \frac { P(\theta_{1} | H^{g_{i}}_{d_{j}} ) P(\theta_{2} | H^{g_{i}}_{d_{j}} ) P(H^{g_{i}}_{d_{j}} )} { \sum_{k=1}^{m \times n}P(\theta_{1}|H_{k})P(\theta_{2}|H_{k})P(H_{k})},
\end{equation}

The generalization of this expression for $f$ observables is:
\begin{equation} \label{equ1}
P(H^{g_{i}}_{d_{j}} | \theta_{1} \cap \theta_{2} \cap ... \cap \theta_{f}) = \frac{P ( H^{g_{i}}_{d_{j}}) \prod_{l=1}^{f} P ( \theta _{l}|H^{g_{i}}_{d_{j}} ) }{ \sum_{k=1}^{m \times n} P(H_{k}) \prod_{l=1}^{f} P(\theta_{l}|H_{k})  }
\end{equation}

\subsection{Practical application of the formalism} 

Now consider the application of this formalism to the specific observables 
of our problem, starting with the apparent $J$ magnitude.
Given an association and a distance, the expected absolute $J$ magnitude 
($\bar\theta$) is deduced from the $I_{c}-J$ color of the candidate 
and the empirical sequence ($M_J$ vs $I_{c}-J$) of the association.
The parameter $\sigma$ represents the dispersion in magnitude for 
a given color index.

Next consider the amplitude of the proper motion as a second observable.
For a given association and position of a candidate on the sky, the 
kinematic model discussed in section~\ref{chap:quatre} predicts the expected 
mean tangential velocity and its dispersion from
the mean and dispersion of $UVW$ of a group (section~\ref{chap:quatre}).
Given this tangential velocity and the distance considered, 
the amplitude of the proper motion ($\bar\theta$) is calculated along 
with its dispersion ($\sigma$).
These quantities are compared to the observed proper motion using Eq.~(\ref{pabaye}).
We apply this methodology for the amplitude of the proper motion in both
right ascension and in declination.

The last two observables to consider are the right ascension and declination, 
$\theta$ = $(\alpha, \delta)$. 
Given the position $(\alpha, \delta)$ of a candidate and a distance, 
the corresponding galactic positions $XYZ$ of the candidate are determined
and compared to the mean ($\bar\theta$) and dispersion ($\sigma$) from 
the kinematic model.

All the parameters used in the analysis are given in Table~\ref{tab:prop2}.
In practice, the analysis was carried out for distances ranging from
1 to 200 pc, by increment of 0.5 pc.
To handle the impact of biased photometry from possibly unresolved binary stars, 
we added an extra hypothesis for each group wherein the photometric sequences 
was shifted by 0.75 mag, thus correcting for equal luminosity binarity. 
In what follows, the probabilities reported for a given group are the sum of the probabilities 
for the nominal and shifted photometric sequence hypotheses. 
Finally, in calculating Eq.~\ref{equ1}, we raised the probabilities for the three galactic positions ($X,Y,Z$) to the 
power 2/3 to ensure that all observables carry an equal weight.
A web-based tool of the Bayesian technique presented here is available at 
\url{http://www.astro.umontreal.ca/\%7emalo/banyan.php}. 
For simplicity, the online version does not include the photometry observables.

\begin{figure}[!h]
\epsscale{1.2}
\plotone{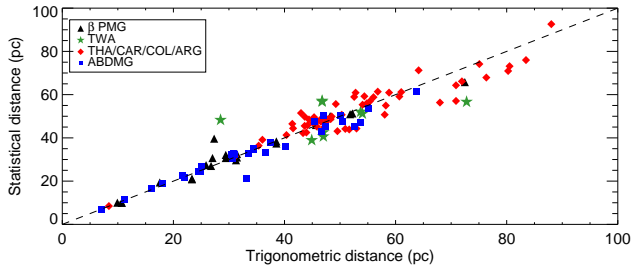}
\caption{\footnotesize{Comparison between the statistical distance from 
Bayesian analysis and the trigonometric distance for the known members of young 
associations.} \label{fig:fig5}}
\end{figure}

To illustrate the robustness of the Bayesian analysis, this 
method was applied to known members of the young associations considered. 
Adopting a membership probability threshold of 90\%, 72\% of the {\em bona fide members}
are recovered (see Table~\ref{tab:allmemb}). 

Figure~\ref{fig:fig5} presents a comparison between the  
observed trigonometric distance ($d_p$) and the \textit{statistical} distance ($d_s$) estimated 
by the Bayesian analysis. 
The statistical distances agree with the trigonometric distances 
within 10\%.

We applied this method to the sample of field stars from \citet{2003phanbao}, and
adopting the same membership threshold (P$_{field} >$ 90\%), 98\% of these 
stars are found to be member of the field. 

It should be stressed that the method above does not make use of the 
radial velocity as an input observable. 
Table~\ref{tab:allmemb} shows the resulting probabilities when the radial velocity is included in the analysis. 
In general, as expected, the probability slightly increases with this 
additional knowledge but this is not always the case;
this will be discussed later in section~\ref{chap:neuf}.

%%%%%%%%%%%%%%%%%%%%%%%%%%%%%%%%%%%%%%%%%%%%%%%%%%%%%%%%%%%%%%%%%%%%%%%%%%%%%%%%%%%%%%%%%%%%%%%%%%%%%%%%%%%%%%%%%%%%%%%
\section{LOW-MASS STAR SEARCH SAMPLE} \label{chap:six}
%%%%%%%%%%%%%%%%%%%%%%%%%%%%%%%%%%%%%%%%%%%%%%%%%%%%%%%%%%%%%%%%%%%%%%%%%%%%%%%%%%%%%%%%%%%%%%%%%%%%%%%%%%%%%%%%%%%%%%%

For our search for new late-type members of young associations, 
we wanted to start with a sample of low-mass stars showing chromospheric 
X-ray and H$\alpha$ emissions, 
which are both indicator of youth.
We used the sample of \citet{2006riaz}, consisting of 1061 spectroscopically 
confirmed K5V to M5V dwarfs, supplemented with 43 K5V-M5V young star 
candidates previously identified by \citet{1997kastner}, \citet{1999webb}, \citet{2000torres},
\citet{2004zuckerman}, \citet{2006torres}, 
\citet{2008torres}, \citet{2009lepine}, \citet{2010looper}, \citet{2010schlieder}, 
\citet{2011kiss}, \citet{2011rodriguez}, \citet{2012schlieder} and \citet{2012bowler}. 

The $I$-band photometric data come 
from the DENIS and SDSS-DR8 catalogs and several other studies 
\citep[][ Riedel (in prep)]{2006torres,2002reid,2010koen,2008casagrande,2003reid,2004reid}.

We have removed stars with $I_{c}$ magnitude uncertainties larger than 0.2 mag
from our search sample.
Also photometric observations of 7 stars with uncertain/missing $I$-band measurements 
were obtained on
2010 August 23-25 (NOAO-2010B-0449) at the CTIO 0.9m telescope using the full 
(13.7\,$\arcmin \times$ 13.7\,$\arcmin$) Tek 2048 $\times$ 2046 CCD camera 
with 0.401\,$\arcsec$ pixel$^{-1}$.
The observations were made using a Gunn-$i$ filter (8200/1500).
Photometric calibration was done using observations of fields containing severals stars with 
known magnitudes and available spectroscopy from SDSS.
Since the SDSS-i filter is not exactly the same as the Gunn-$i$ filter, 
synthetic magnitudes were extracted from the spectra of the SDSS stars.
One standard field was observed after each target observation.
The $I_{c}$ magnitudes are given in Table~\ref{tab:candprop} (see note e).
Finally, the $J$-band data are taken from the 2MASS PSC catalog.

The proper motion data mainly come from the NOMAD \citep{2005zacharias}, 
UCAC3 \citep{2009zacharias}, PPMXL \citep{2010Roeser} and
other catalogs. 
From the original sample of 1104 stars, we kept 758 stars with proper 
motion measured at a significance of more than four sigma and a 
good $I_{c}$ magnitude measurement. 
Of these 758 stars, 71 K5V-M5V were previously identified as 
young stars in the literature. 

%%%%%%%%%%%%%%%%%%%%%%%%%%%%%%%%%%%%%%%%%%%%%%%%%%%%%%%%%%%%%%%%%%%%%%%%%%%%%%%%%%%%%%%%%%%%%%%%%%%%%%%%%%%%%%%%%%%%%%%
\section{RESULTS} \label{chap:sept}
%%%%%%%%%%%%%%%%%%%%%%%%%%%%%%%%%%%%%%%%%%%%%%%%%%%%%%%%%%%%%%%%%%%%%%%%%%%%%%%%%%%%%%%%%%%%%%%%%%%%%%%%%%%%%%%%%%%%%%%

\subsection{Identification of new candidates}

\begin{figure}[!hpb]
\epsscale{1.2}
\plotone{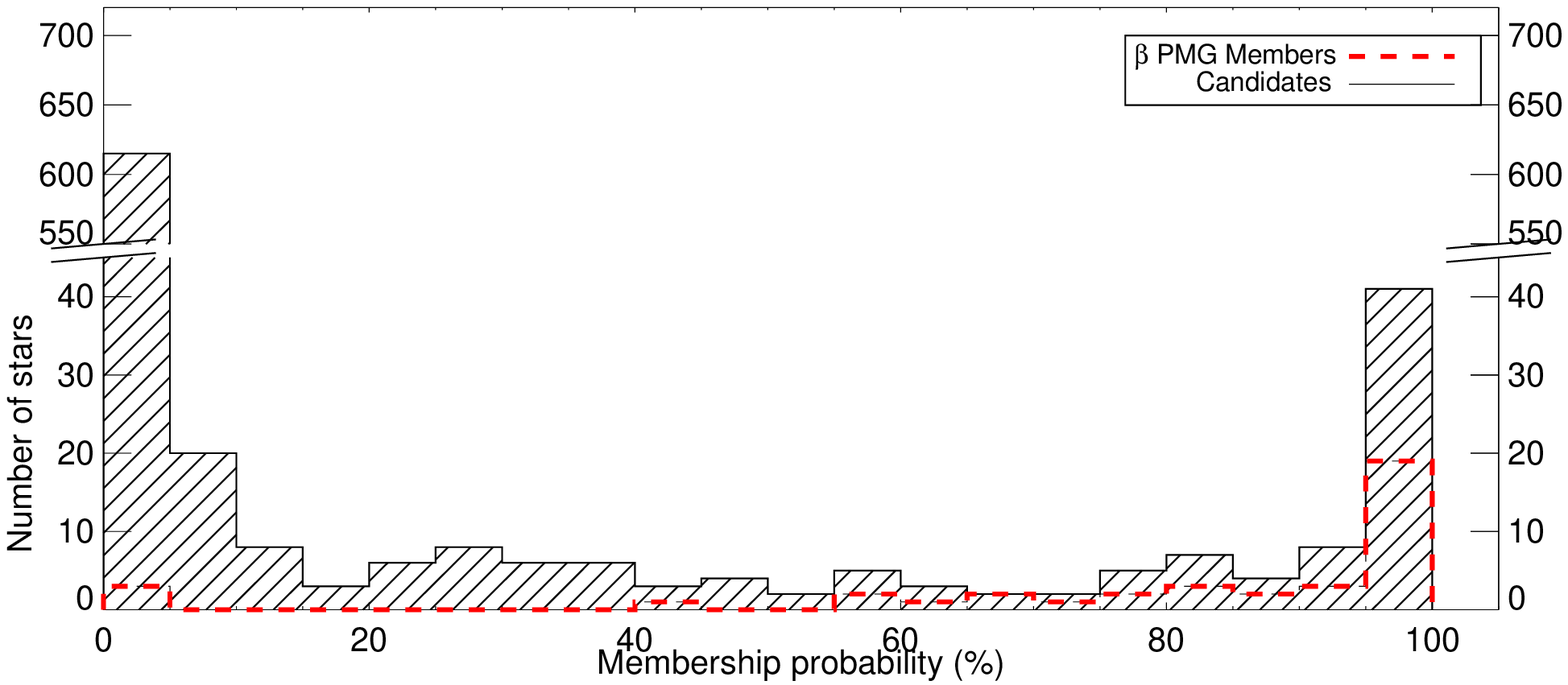}
\\[-2.0ex]
\epsscale{1.2}
\plotone{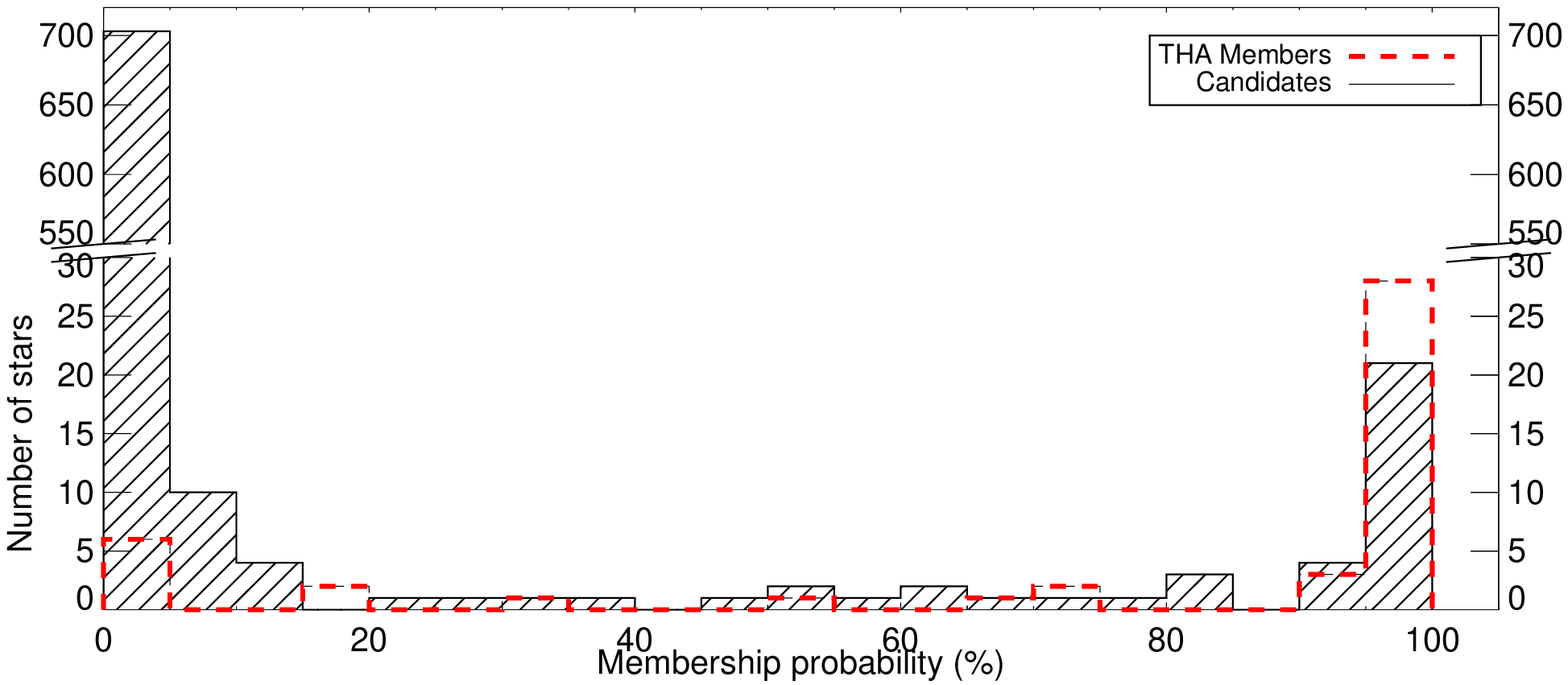}
\\[-2.0ex]
\epsscale{1.2}
\plotone{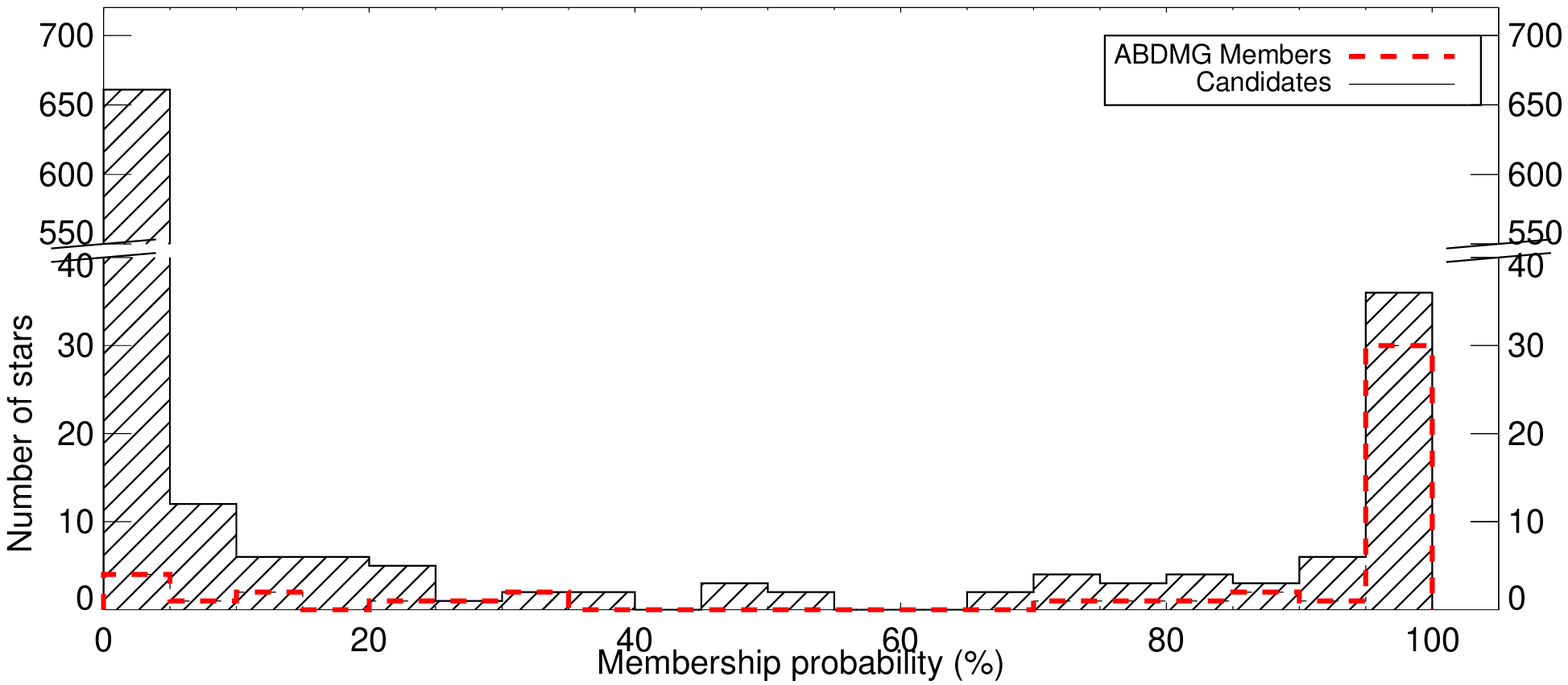}
\caption{\footnotesize{Distribution of membership probability for the 
entire search sample for each young association (black histogram). 
The {\em bona fide members}
are shown with a red dashed histogram. 
The vast majority of the 758 stars have probabilities close to 0\% as one expects.}
\label{fig:fig6}}
\end{figure}

As described in the last section, our initial search sample is 
subject to the Bayesian analysis using as observables 
the apparent $I_{c}$ and $J$ magnitudes, the amplitude of proper motion in
right ascension and declination, and the right ascension and declination.
We will see later (section~\ref{chap:huit}) how other observables, in 
particular the radial velocity, can be used to better constrain the 
membership probability of a candidate.
Figure~\ref{fig:fig6} presents, for the entire search sample, 
the membership probability distribution for three of the seven young 
associations considered.
As expected, the vast majority of stars have very low membership 
probabilities, but there is a significant population of objects with a
probability higher than 90\%. We set a threshold at 90\% 
to qualify a candidate as a ``high probability member'' of a given 
association (see next section for a discussion of false positives). 
Overall, our analysis has identified a total of 215 highly probable 
members including 58 young candidate identified in the literature.

The sample of new candidates (215) from our study 
includes one star in TWA, 37 in $\beta$PMG, 17 in THA, 20 in COL, 6 in CAR, 
50 in ARG and 33 in ABDMG. 
The properties of all candidates are presented in Table~\ref{tab:candprop}. 
We note that 51 of our candidates have an ambiguous membership, i.e., 
they are probable candidates in more than one association with a combined
probability over all young groups above 90\%. 
A radial velocity measurement is needed to remove this ambiguity 
(see Tables~\ref{tab:candall},~\ref{tab:allcand}). 

\begin{figure}[!phb]
\epsscale{1.1}
\plotone{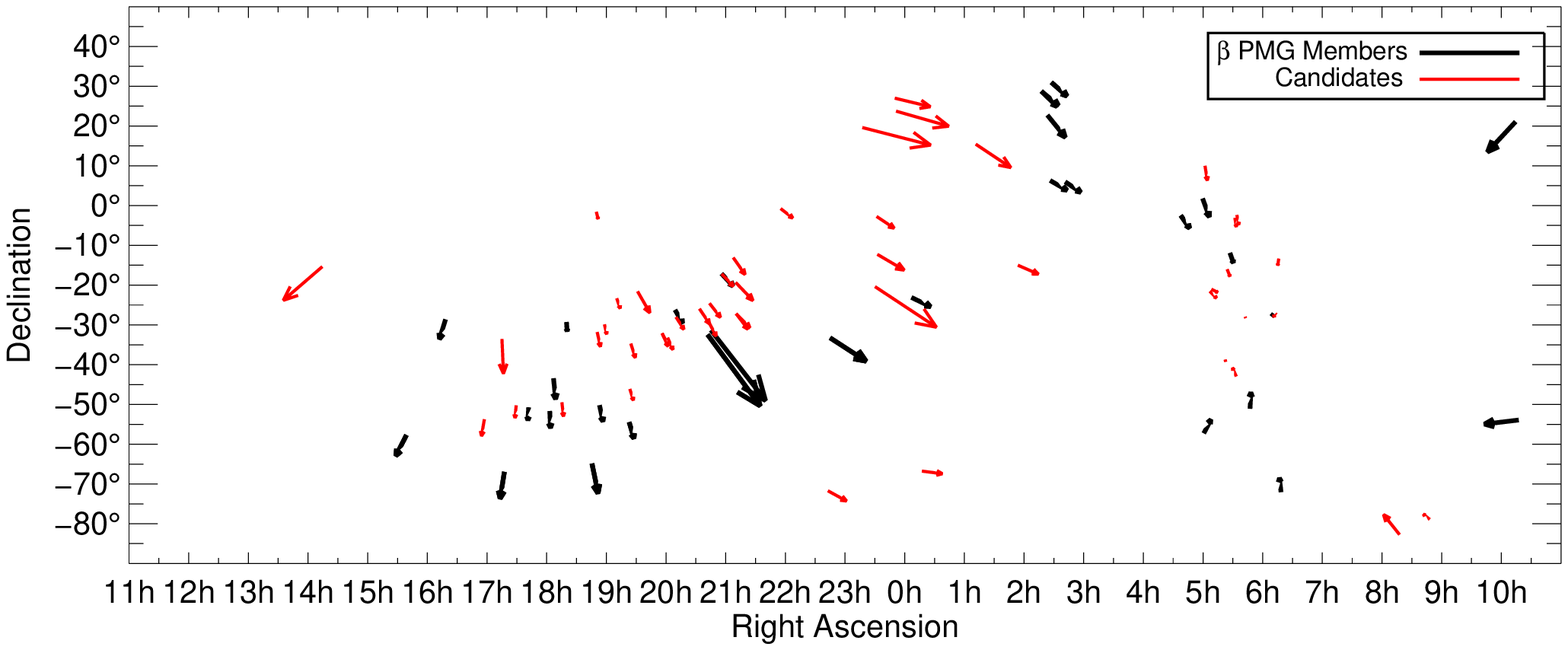}
\\[-2.0ex]
\epsscale{1.1}
\plotone{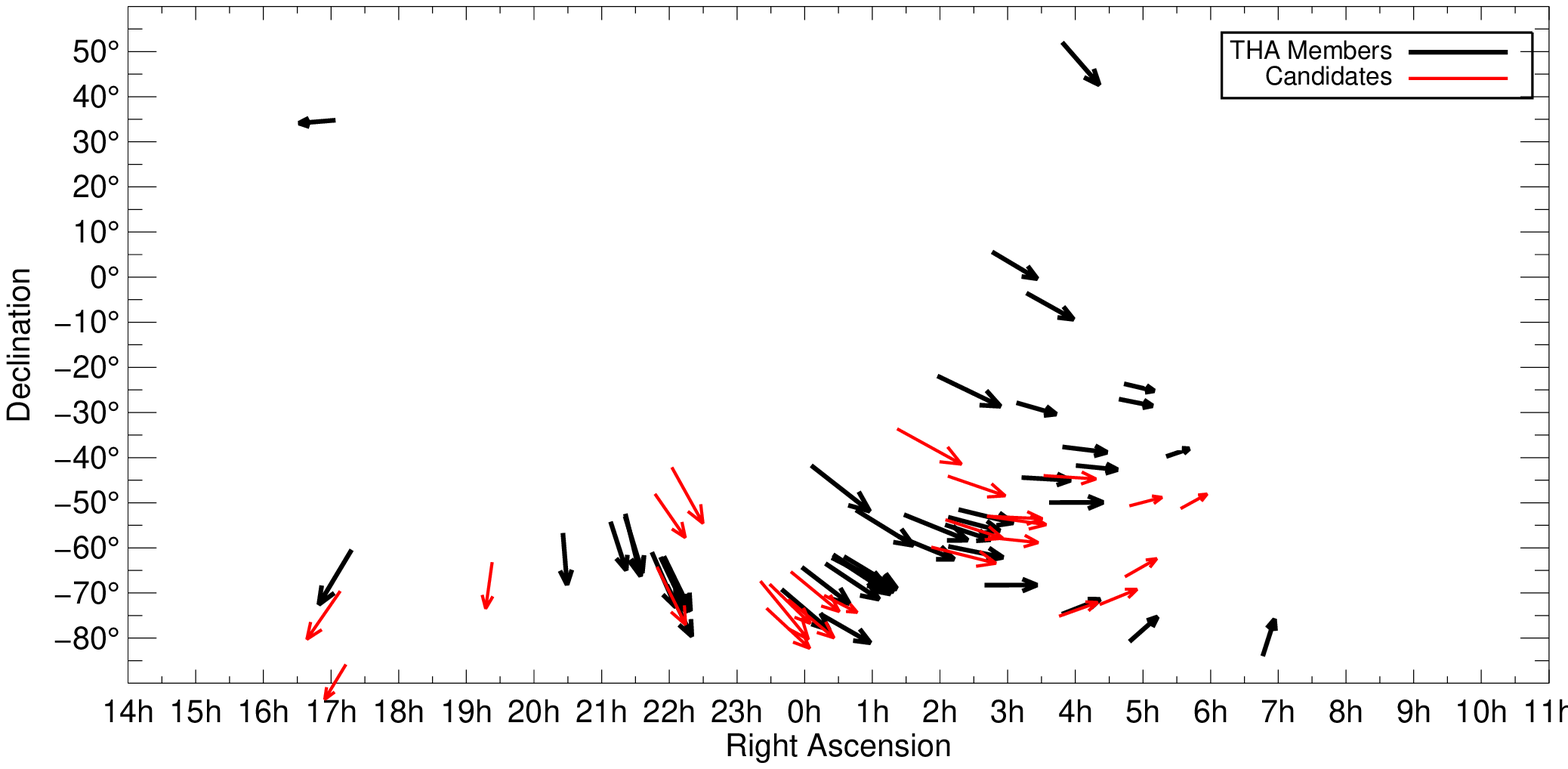}
\\[-2.0ex]
\epsscale{1.1}
\plotone{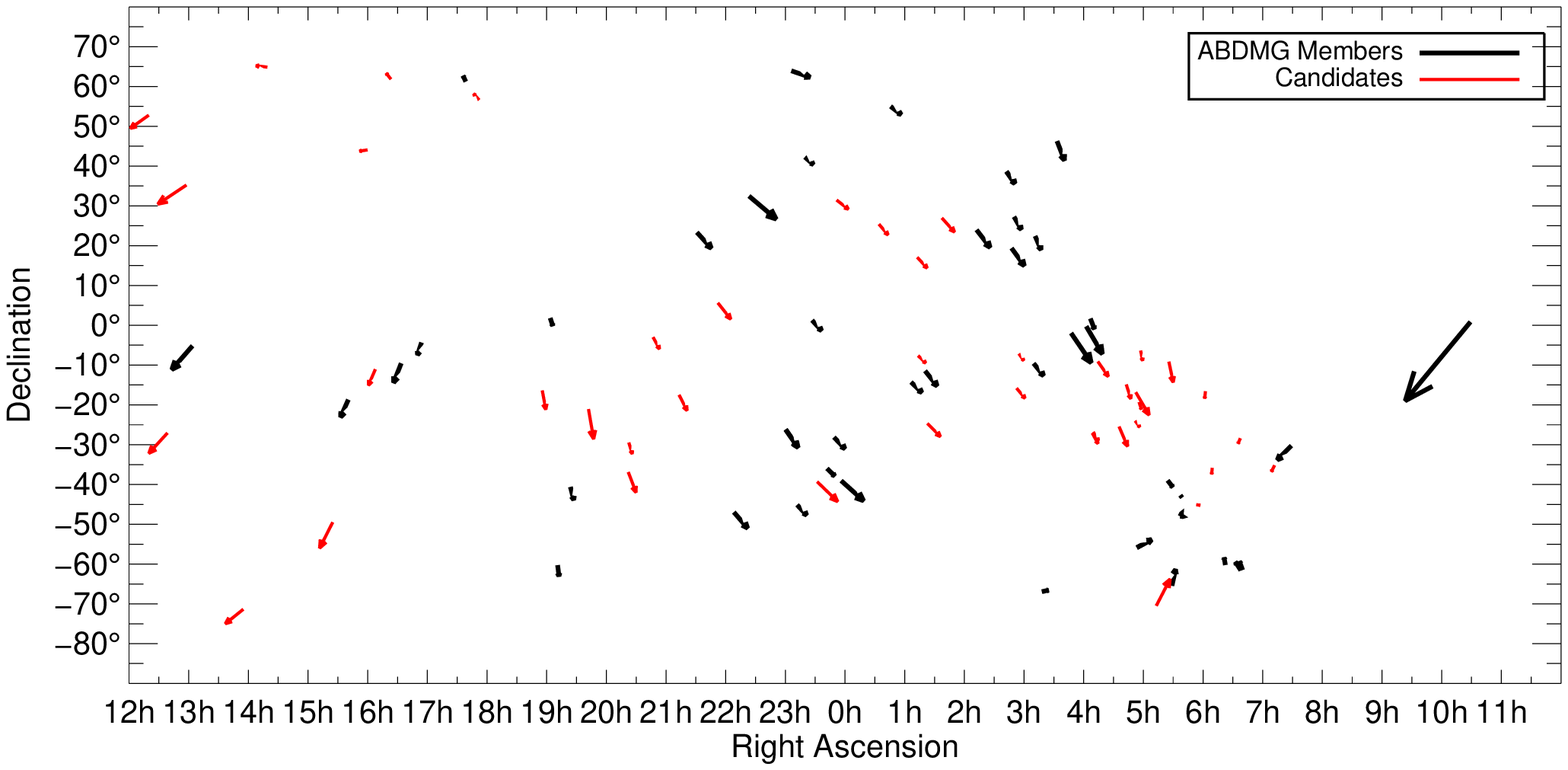}
\caption{ \footnotesize{Position on the sky and vector of proper motion 
for the {\em bona fide members} (bold black arrows) and the new candidates (red arrows) 
resulting from this study.}
\label{fig:fig7}}
\end{figure}

Figure~\ref{fig:fig7} presents the position on 
the sky as well as the amplitude and the direction of the 
proper motion for both {\em bona fide members} and new 
highly probable candidate 
members of $\beta$PMG, THA and ABDMG.
The overall trend in amplitude and direction of the proper motion for the 
candidate members agree well with the {\em bona fide members}.
Although the majority of the candidates (164) are located in the 
Southern hemisphere, 13 candidates are in the Northern hemisphere: 4 
in $\beta$PMG, 4 in ARG and 5 in ABDMG.

% PROBABILITY DISTRIBUTION

% POSITION ON SKY

% SPECTRAL TYPE
\begin{figure}[!t]
\epsscale{1.2}
\plotone{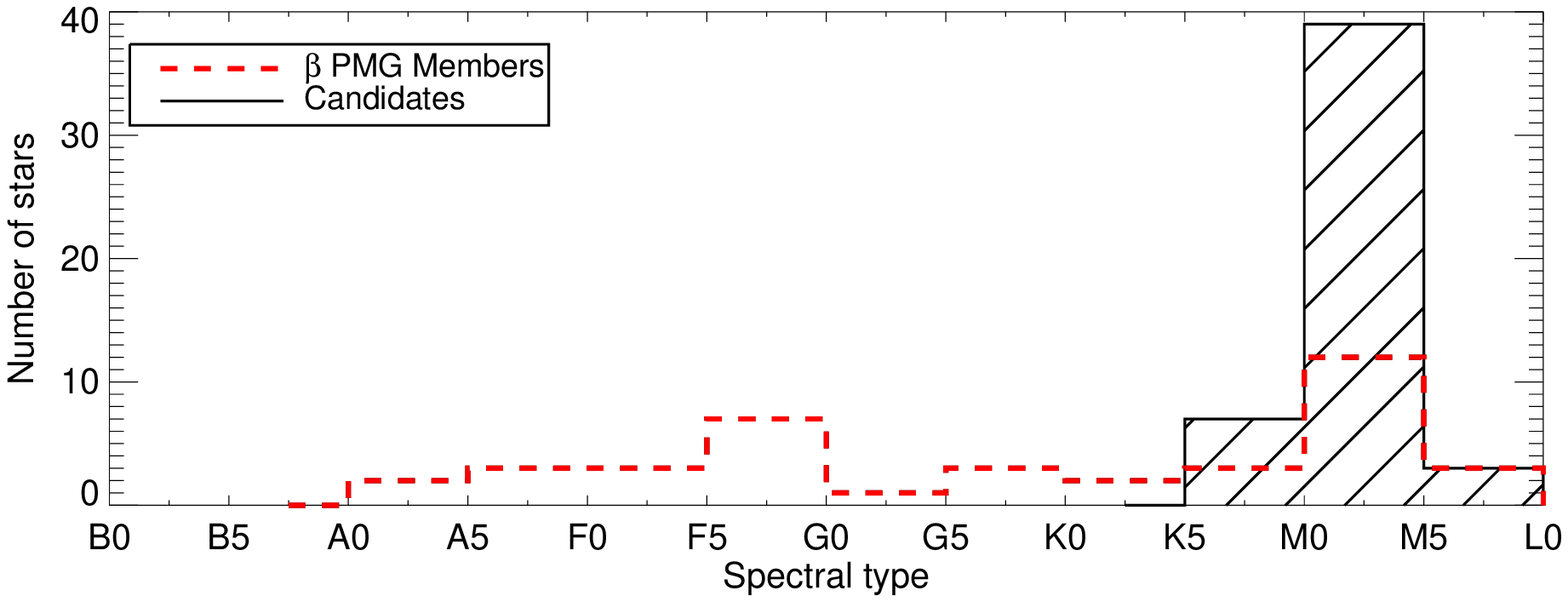}
\\[-5.0ex]
\epsscale{1.2}
\plotone{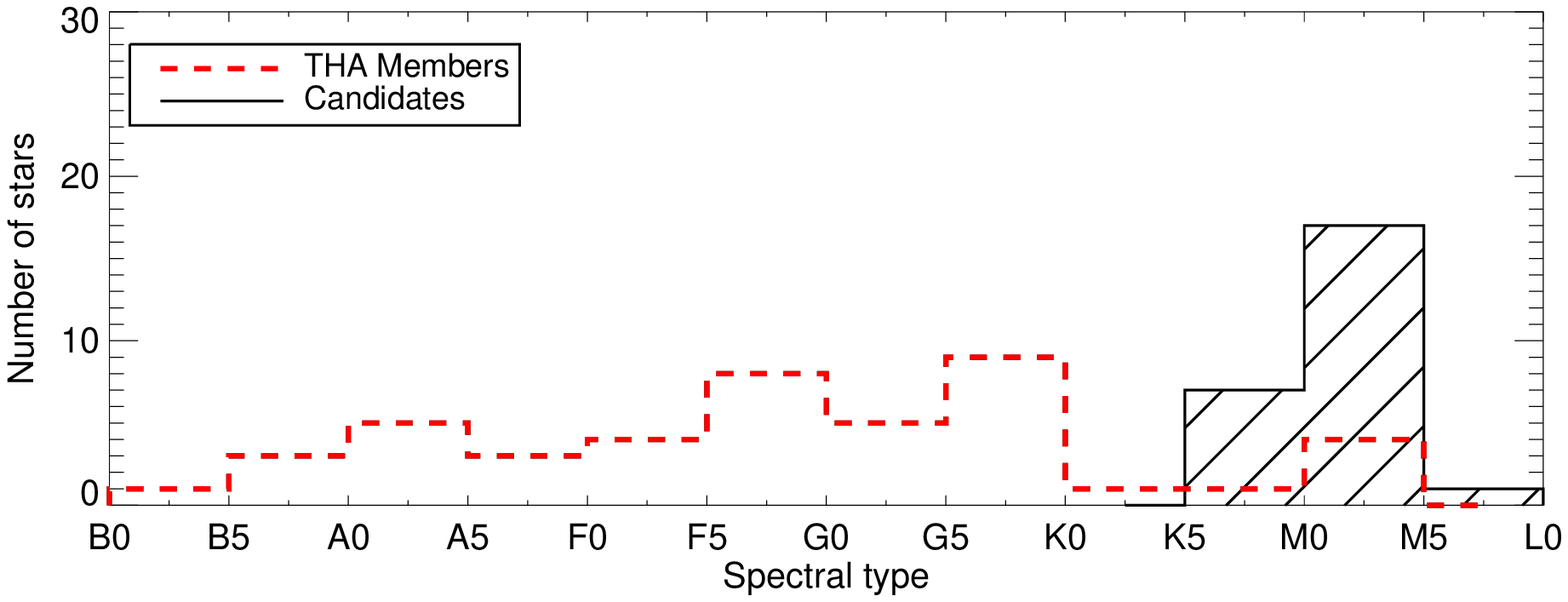}
\\[-5.0ex]
\epsscale{1.2}
\plotone{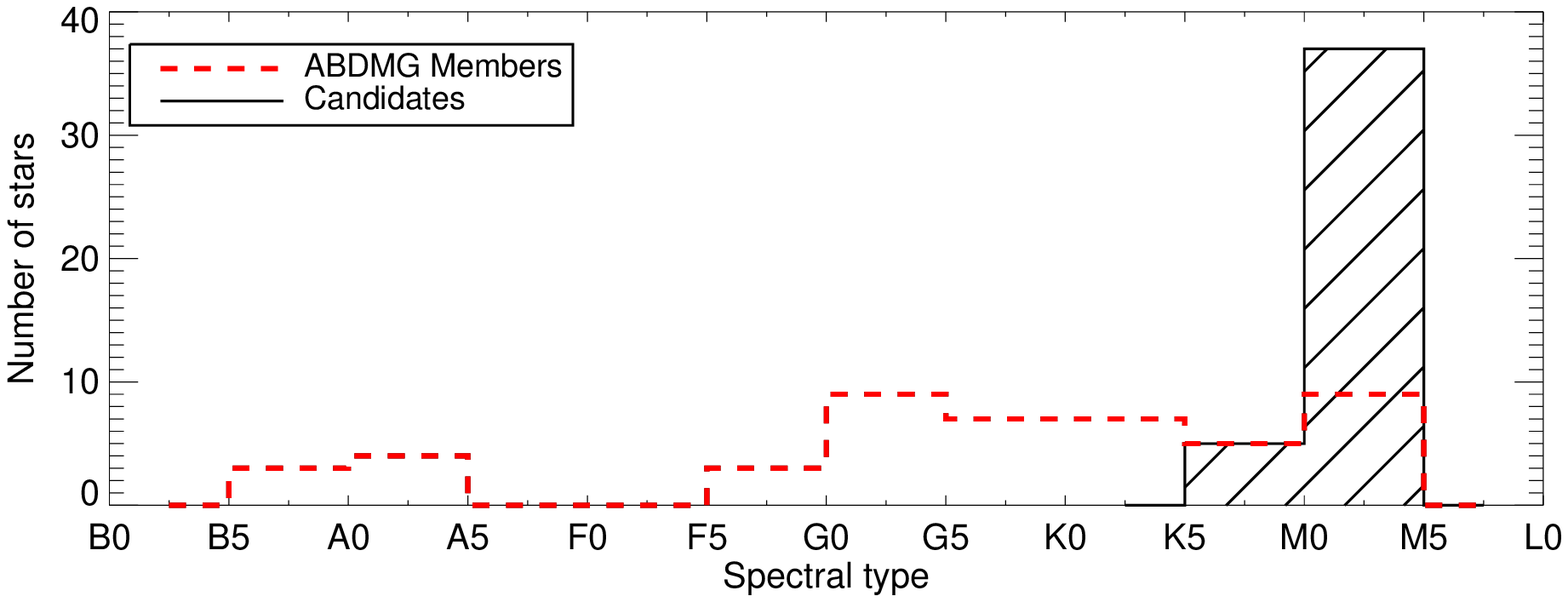}
\caption{\footnotesize{Spectral type distribution of previously known 
members (red dashed lines) and candidates
(bold black lines) from this work.}
\label{fig:fig8}}
\end{figure}

% DISTANCE DISTRIBUTION
\begin{figure}[!t]
\epsscale{1.2}
  \plotone{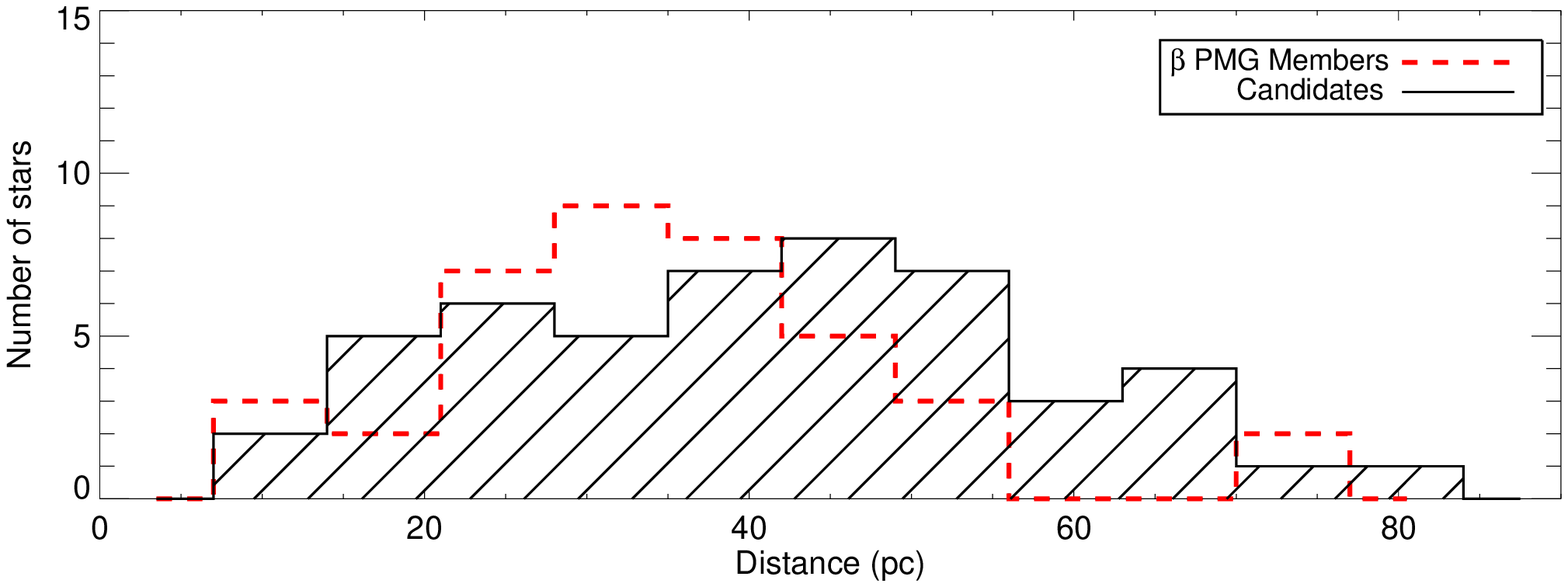}
\\[-4.5ex]
\epsscale{1.2}
  \plotone{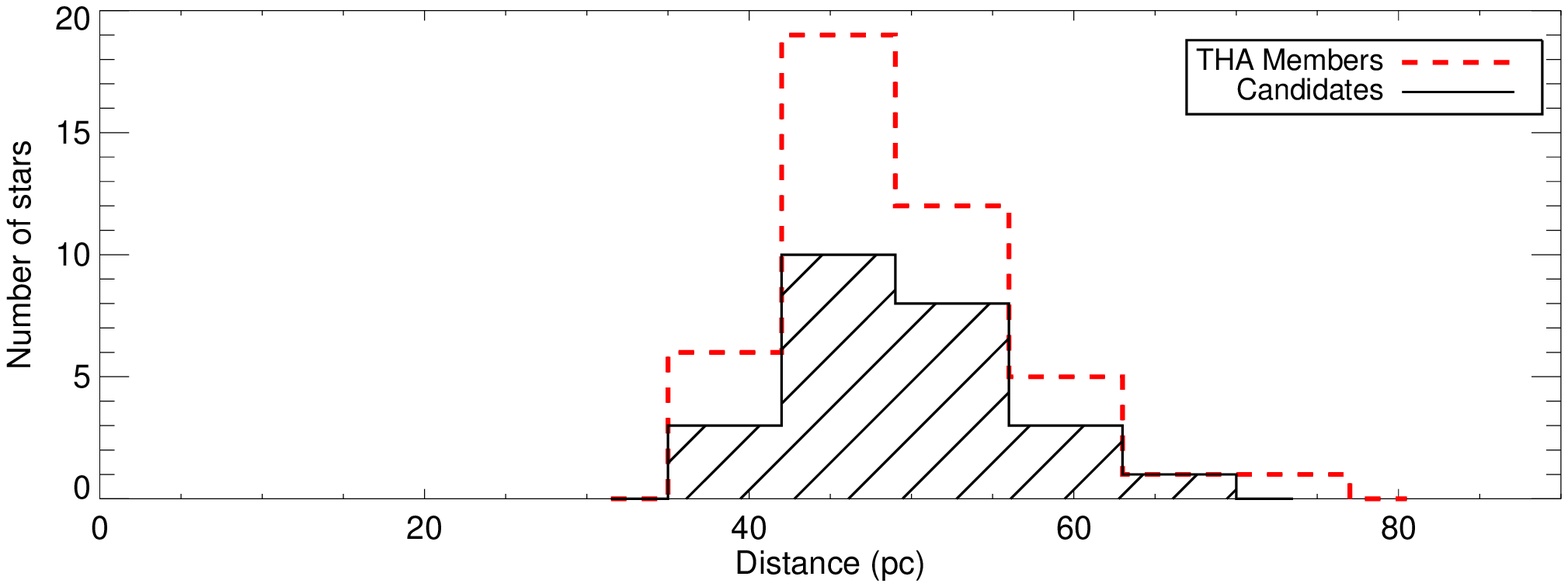}
\\[-4.5ex]
\epsscale{1.2}
  \plotone{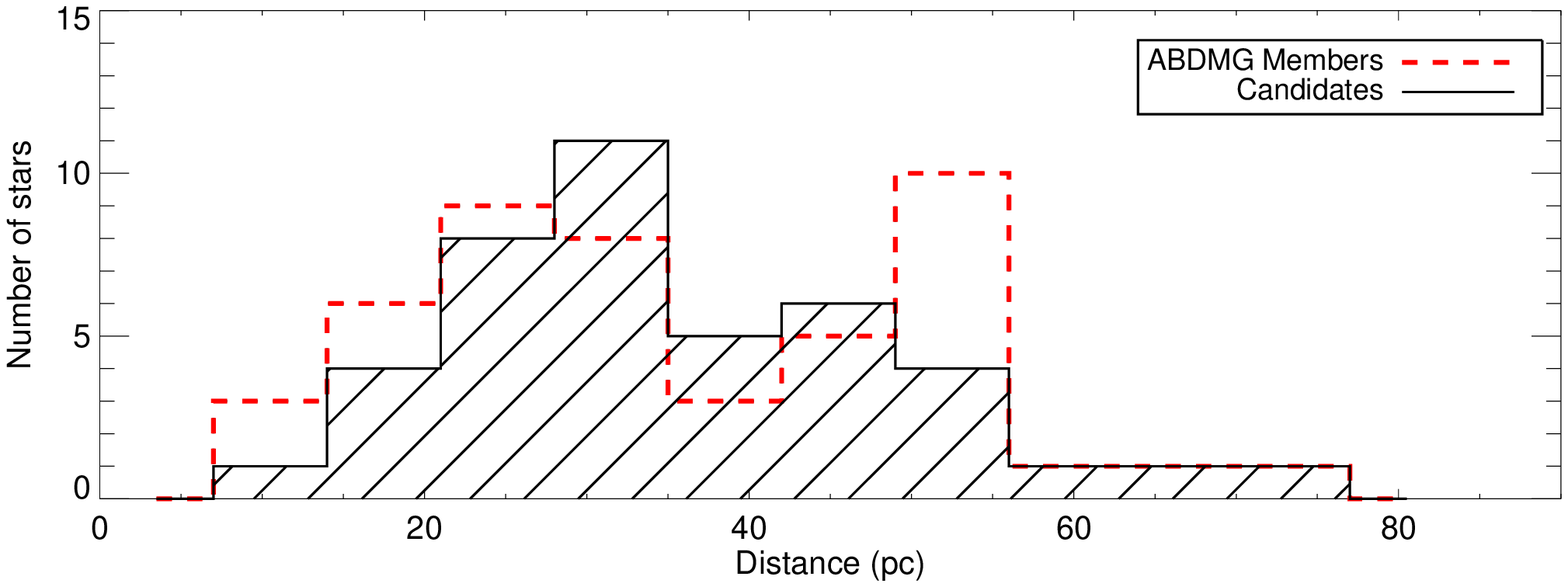}
\caption{\footnotesize{Distance distribution for the {\em bona fide members}
(red dashed lines) and for candidates (bold black lines) 
from this study.} \label{fig:fig9}}
\end{figure}

% XYZ PLOT
\begin{figure}[!b]
\epsscale{1.2}
  \plotone{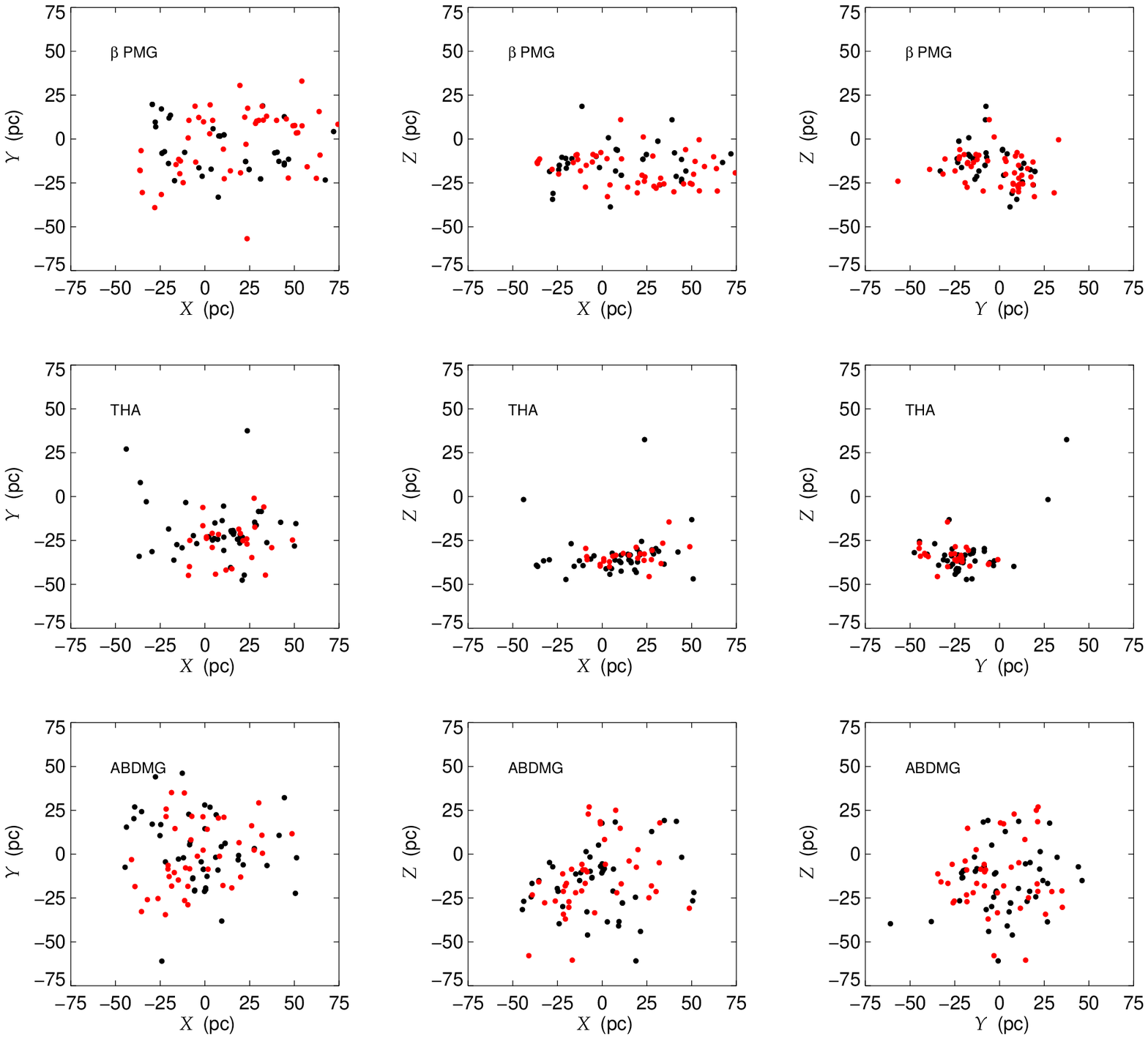}
\caption{\footnotesize{Galactic position $XYZ$ of the {\em bona fide members}
(black circles) and new candidates (red circles) from this study.}
\label{fig:fig10}}
\end{figure}

Figure~\ref{fig:fig8} shows the spectral type distribution of 
the new candidates compared to the {em bona fide members} of 
$\beta$PMG, THA and ABMDG.
All the distributions show a maximum between spectral types 
M0V and M5V. 
Excluding ambiguous members, our analysis have unveiled 164 new late 
type (K5V-M5V) candidates.
Compared to the 43 {\em bona fide} K5V-M5V dwarfs already catalogued in the 
seven kinematic groups under study here, this new sample of young M dwarfs, 
if confirmed, would roughly quadruple the known population of late type dwarfs 
of these associations.

The distance distributions for the {\em bona fide members} of $\beta$PMG, THA and ABMDG 
are superimposed to those of the candidates in figure~\ref{fig:fig9}. 
Overall, the distance distributions for both {\em bona fide members} 
and candidates are very similar. 
For ABDMG, our analysis finds a tail of a few candidates between 
55 and 80\,pc.
Our analysis also identifies a few candidates that are
relatively close ($<$20\,pc) to the Sun.
In general, candidates found at relatively large distances should be taken with caution
as they may well be members of more distant ($>$100\,pc) young co-moving groups not
considered in our analysis.
One such example is the 8\,Myr $\epsilon$Cha association which could be easily confused
with $\beta$PMG since both associations have very similar $UVW$ \citep{2008torres}.
This emphasizes the need for a trigonometric parallax as a necessary criterion for confirming 
the {\em bona fide} status of a candidate.

The galactic positions $XYZ$ of the candidates are presented in 
figure~\ref{fig:fig10}. The dispersion of the {\em bona fide members} and the 
new candidates are very similar, in particular for $Y$ and $Z$. 
This is expected since the candidates were chosen to be within the parameters of
the {\em bona fide members}.

\subsection{Quantifying the contamination} \label{sect:septdeux}

It is expected, that some field stars will have a high
membership probability by pure chance. 
One thus needs to estimate the number of such contaminant in our sample.
To quantify the number of false positives, we repeated our analysis 200 times
by considering, each time, seven fake young
associations with $UVW$ and $XYZ$ distributions similar to, but not 
overlapping with, those of the seven young 
associations studied here. 
The seven fake associations were divided into three representative age 
groups: group 1 ( 8-20 Myr; BPMG, TWA), group 2 (20-50 Myr; COL, THA, CAR, ARG) 
and group 3 (>50\,Myr; ABDMG). 
As before, the candidate membership probability threshold was set to 90\%.

\begin{figure}[!bth]
\epsscale{1.2}
\plotone{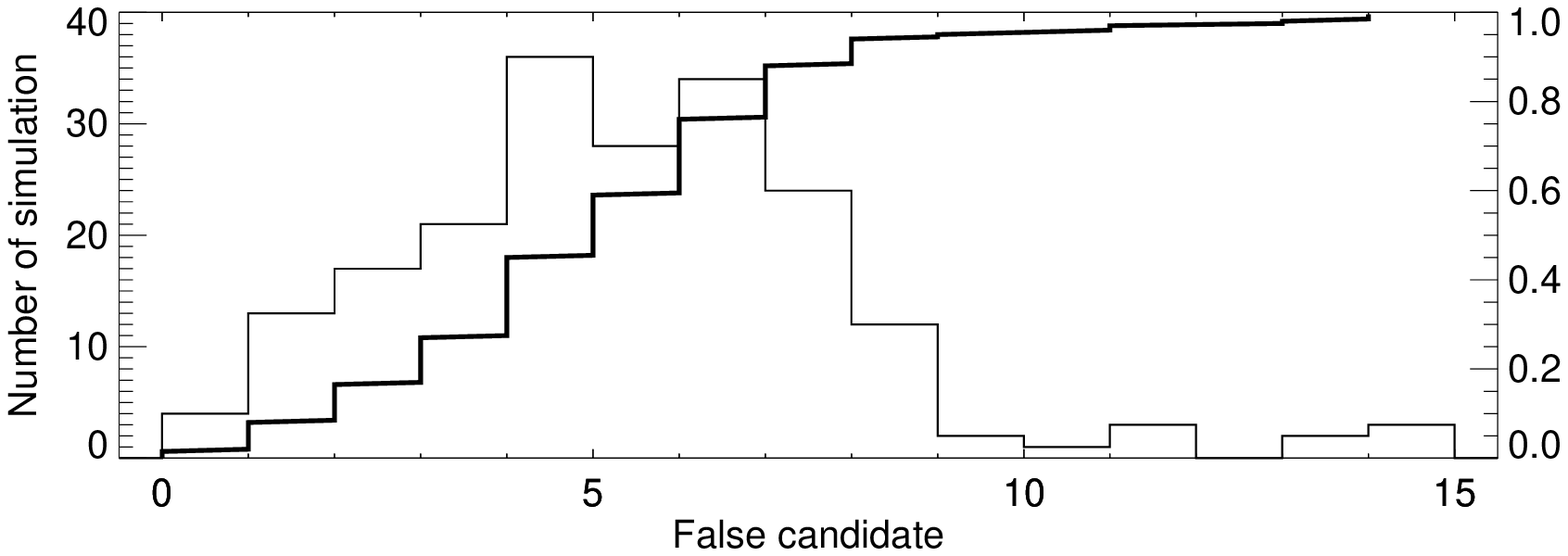}
\plotone{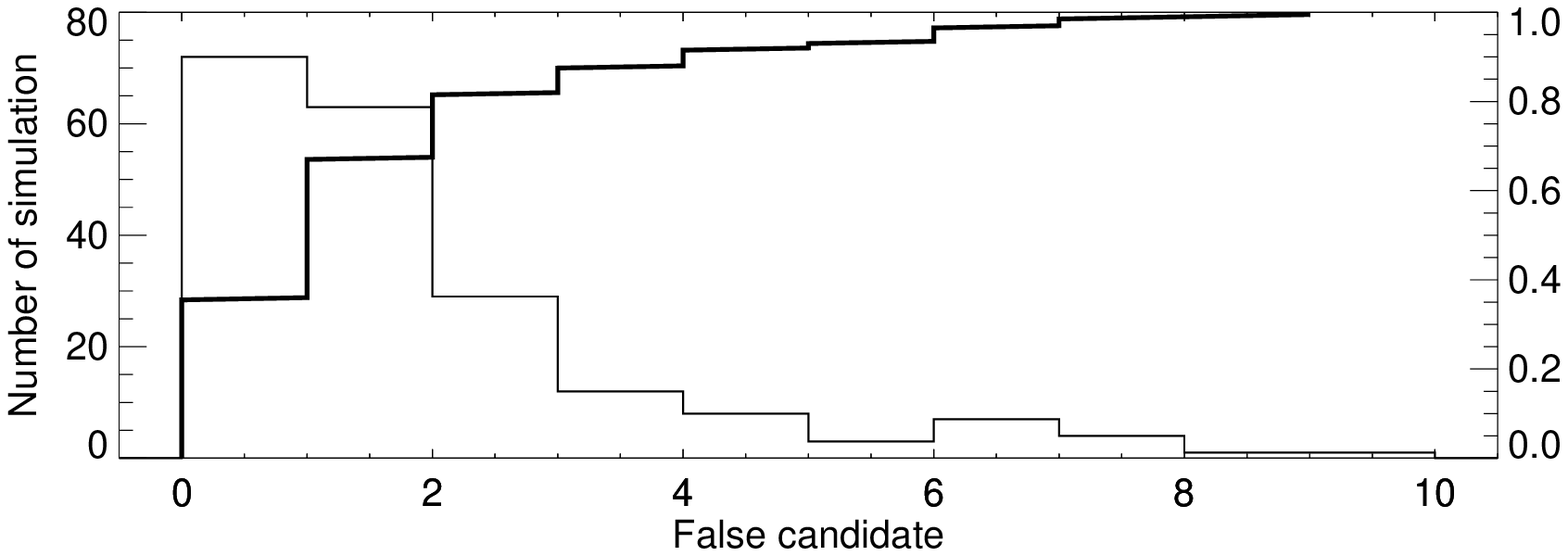}
\plotone{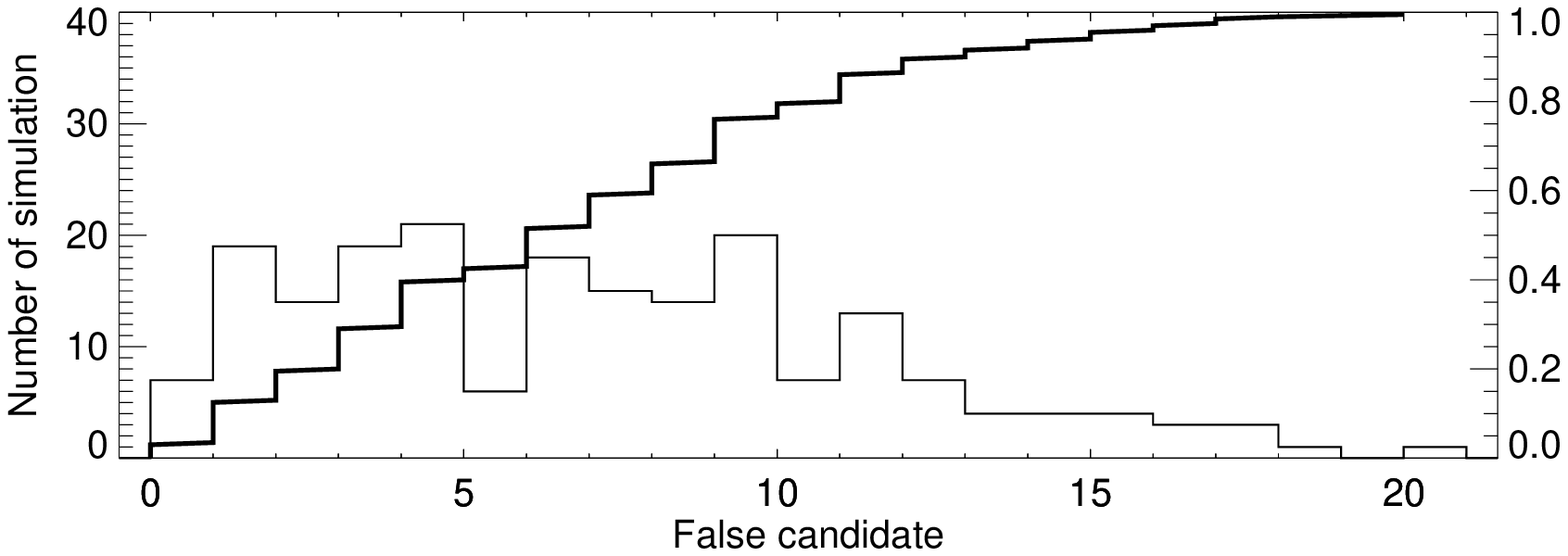}
\caption{\footnotesize{Distribution of number of candidates in fake 
associations (thin line and left axis) for various 
age groups: 8-20\,Myr (top), 20-50\,Myr (middle) and >50\,Myr (bottom). 
The thick solid line is the corresponding cumulative distribution  (right axis). } \label{fig:fictive}}
\end{figure}

These Monte Carlo simulations allow determining the median number of false positives. %with a 1$\sigma$ (68\%). 
The results of this procedure are presented in figure~\ref{fig:fictive}. 
The contamination is relatively modest for group 1 and 2 with typically (median) 
only a few false positives, which 
is about an order of magnitude less than the real number of candidates. 
For false associations in the group 1 (8-20\,Myr), 
the number of expected false positives is also smaller than
 the number of candidates by a factor of 10. 
The number of contaminants is significantly larger for ABDMG (group 3), as 
one would expect given that the CMD sequence 
for the older ( >50\,Myr) association
lies closer to that of field stars compared to younger sequences.
Based on this analysis, we can estimate a typical (median) false alarm rate 
of $\sim$ 5\% for group 1, 
$\sim$ 10\% for group 2 and $\sim$ 14\% for group 3, suggesting that a large fraction of 
our candidates are likely to be genuine members.

% FALSE CANDIDATES

%%%%%%%%%%%%%%%%%%%%%%%%%%%%%%%%%%%%%%%%%%%%%%%%%%%%%%%%%%%%%%%%%%%%%%%%%%%%%%%%%%%%%%%%%%%%%%%%%%%%%%%%%%%%%%%%%%%%%%%
\section{Radial velocity and lithium follow-up} \label{chap:huit}
%%%%%%%%%%%%%%%%%%%%%%%%%%%%%%%%%%%%%%%%%%%%%%%%%%%%%%%%%%%%%%%%%%%%%%%%%%%%%%%%%%%%%%%%%%%%%%%%%%%%%%%%%%%%%%%%%%%%%%%

As pointed out above, the statistical analysis
can be performed with other observables, for example the radial velocity (RV).
The previous analysis was performed without prior knowledge of this measurement. 
It is of course desirable to add this observable to the 
Bayesian analysis if the information is available. 

Furthermore, while the kinematics of these stars indicate membership in these 
associations, their youth should be confirmed through the measurement of 
other spectroscopic age indicators such as lithium (Li) absorption at 
6707.8 \AA\,, a youth indicator for K-M dwarfs less than 70 Myr 
\citep[see fig.5; ][]{2008mentuch}.
This section presents (ongoing) follow-up observations (RV and Li) of a subset of 
the highly probable 
members identified by our analysis.

\subsection{Radial velocity follow-up}

We initiated a program to obtain near-infrared high-resolution spectroscopy 
of some of our highly probable members to measure their radial velocities. 
The observations were performed using the PHOENIX \citep{2003hinkle} 
spectrograph at the Gemini South telescope (GS-2009A-Q-89, GS-2009B-Q-45, GS-2010A-Q-32, GS-2010B-Q-18, GS-2010B-Q-89). 
We used a $0.34\arcsec$-wide slit in combination with the H6420 filter 
(1.547$\mu$m-1.568$\mu$m) for a resolving power of R$\sim\,30 000$. 
The instrument setup was inspired by the work of \citet{2002mazeh} on low-mass binaries. 
The observations were obtained with a typical ABBA dither pattern along the 
slit with individual exposures of 60 to 300\,s depending on the target brightness. 
Flat field and dark images were used to correct for detector cosmetics and OH 
night-sky emission lines were used for wavelength calibration. 
A set of radial velocity standards were observed and served as cross-correlation 
spectra for radial-velocity measurements. 

To date, we have obtained RV measurements at one epoch for 101 candidates. 
The results show that 50\% of them are fast rotators which, by itself, is an 
indication of youth, but makes the measurement of a radial velocity difficult. 
For 14 of the slow rotators we have obtained a second epoch of RV 
measurement (to assess spectroscopic binarity);
these measurements are presented in Table~\ref{tab:candprop}. 
As discussed in section~\ref{chap:neuf}, all 14 slow rotators have radial 
velocities in good agreement with the expected values 
for their respective association.

\subsection{Lithium follow-up}

\begin{figure}[!htb]
\epsscale{1.2}
\plotone{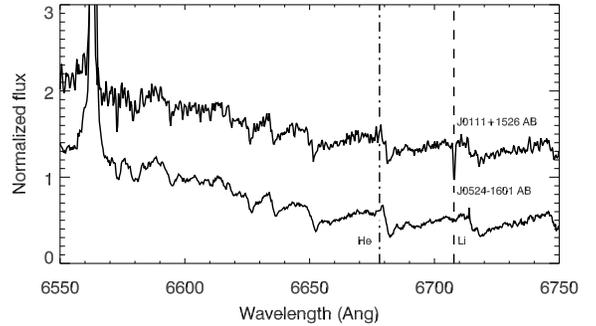}
\caption{ \footnotesize{High-resolution optical ESPaDOnS spectra of two young binary candidate members showing
lithium absorption (unresolved spectra). The Li feature of J0524-1601 is clearly broadened due to relatively fast rotation.}
\label{fig:lithium}}
\end{figure}

As low-mass stars have a convective outer envelope, the primordial lithium 
in their photosphere is, with time, transported to the center of the star, 
where it is quickly destroyed. 
The presence of lithium in the photosphere of a low-mass star is thus 
indicative of youth.
Lithium absorption at 6707.8 \AA\ is indeed seen for young low-mass members 
of $\beta$PMG, THA and ABDMG. 
While our candidates are highly probable new members of these young 
associations, measurements of various age-dating indicators, such as the 
presence of lithium, is valuable to confirm their membership. 
We have initiated a program to obtain these measurements for the candidate 
members with a membership probability $> $90\%.
High-resolution optical spectroscopy was obtained in 
queue service observing (QSO) mode with ESPaDOnS \citep{2006donati} on 
CFHT. 
ESPaDOnS was used in a ``star + sky'' mode combined with the ``slow'' CCD 
readout mode,
to get a resolving power of R $\sim$ 68000 covering
the 3700 to 10500 \AA\ over 40 grating orders. 
The data were reduced by the QSO team using the CFHT pipeline called 
UPENA 1.0. 
This pipeline uses J-F. Donati's software Libre-ESpRIT \citep{1997donati}. 
The total integration time per target is between 30 and 80 minutes.
So far 28 candidates have been observed and we have analysed the spectrum 
for 10 stars, two of which (J0111+1526 and J0524-1601) show clear lithium 
absorption, confirming their youth. The spectra of these two stars are 
shown in Figure~\ref{fig:lithium} and further discussed in 
the next section.
The results of the remaining stars with follow-up spectroscopy will be 
published in a forthcoming paper.

The lithium detection limit is around K7V at an age of
ABDMG, M0V for THA and M3V for $\beta$PMG \citep{2008mentuch}. 
From our 164 candidate members, we expected to detect lithium for 22, 16, 1 stars 
in $\beta$PMG-TWA,THA-COL-CAR-ARG and ABDMG, respectively. 
For star with a mass lower than the limit detection, we need to find a 
better age-dating indicator, such as surface gravity.

%%%%%%%%%%%%%%%%%%%%%%%%%%%%%%%%%%%%%%%%%%%%%%%%%%%%%%%%%%%%%%%%%%%%%%%%%%%%%%%%%%%%%%%%%%%%%%%%%%%%%%%%%%%%%%%%%%%%%%%
\section{Discussion} \label{chap:neuf}
%%%%%%%%%%%%%%%%%%%%%%%%%%%%%%%%%%%%%%%%%%%%%%%%%%%%%%%%%%%%%%%%%%%%%%%%%%%%%%%%%%%%%%%%%%%%%%%%%%%%%%%%%%%%%%%%%%%%%%%

In light of the statistical analysis presented above, we discuss the 
membership of specific stars drawn 
from three samples: {\em bona fide members}, new highly probable low-mass 
star candidates identified for the first time as part of this work, and 
the other 71 young low-mass star candidates previously identified in 
the literature (see section~\ref{chap:six}). 
For the latter sample, only spectral types later than K5V 
are considered; the most massive ones will be the subject of a 
future publication. 

We should note that some {\em bona fide members} show a high membership 
probability when considering the binary hypothesis. 
This hypothesis should be interpreted as an overluminosity compared to other 
young stars, which could be due to the presence of a binary companion, 
chromospheric activity, or peculiar colors of the star.

\subsection{Bona fide members}

Since the observational properties of {\em bona fide members} form the basis 
of our Bayesian tool, one should expect our analysis to 
yield a high membership probability for these stars.  
This is indeed the case. 
Table~\ref{tab:allmemb} gives the membership probablility obtained for 
these stars for three cases: (1) without considering the RV or parallax 
information (P), (2) considering the RV but not the parallax (P$_{v}$), 
and (3) considering both the RV and parallax (P$_{v+\pi}$). 
For all cases, a binary hypothesis is also considered and whenever it 
yields a higher probability than the single hypothesis, this is 
noted in Table~\ref{tab:allmemb}. 
Assuming that a member is recovered when its probablility is higher 
than 90\%, our analysis correctly recovers 72\% (128/177) of all 
{\em bona fide members} for case 1. 
This fraction increases to 88\% (155/177) when RV is included (case 2) and 
to 91\% when both RV and parallax informations are used (case 3). 
This large recovery rate is expected but there remains a few cases 
(16 out of 177) that have ambiguous or uncertain membership. 
Those are discussed below. 

\subsubsection{$\beta$PMG}
{\bf HIP 12545 AB} : 
This star is a K6Ve spectroscopic binary \citep[sb1;][]{2008torres} that 
was proposed by \citet{2004zuckerman} to be a member of $\beta$PMG. 
From our analysis, this star seems to be a member of COL 
(P$_{v}$=99\%, P$_{v+\pi}$=88\%, P$_{\pi}$=48\%).
The Galactic positions ($XZ$) of this star are
much similar to COL {\em bona fide members}. 
Its measured RV (8.25$\pm$0.05 km s$^{-1}$) is closer 
to that predicted for COL (8.6 km s$^{-1}$) compared 
to $\beta$PMG (11.3 km s$^{-1}$ ). 
However, this difference of 2.7 km s$^{-1}$ is marginal, especially 
since this star is a fast rotator 
\citep[vsini=40 km s$^{-1}$;][]{2006torres} as well as a spectroscopic binary; 
its RV may thus be somewhat uncertain. 
Our analysis favors a membership in COL due to its position
on the CMD which is marginally more consistent (by $\sim$ 0.7\,mag) with COL
compared to $\beta$PMG.
However, when the photometry and the RV are excluded, our analysis favors a membership
in $\beta$PMG (P$_{\pi}$=66\%).
We should note that this star has an EW$_{\rm Li}$=433$\pm$23 m\AA\, 
\citep{2008mentuch} that may be too high to be a member of COL for 
an age of 20-40 Myr.
For this reason, HIP 12545 AB is a {\em bona fide member} of $\beta$PMG. 

{\bf HIP 11360} :
This star (F4IV) was proposed by \citet{2006moor} to be 
a member of $\beta$PMG. 
\citet{2008torres} also proposed this star as a candidate member of $\beta$PMG 
since its kinematic is similar to HIP 12545 AB. 
Our analysis suggests this star to be in COL 
(P$_{v}$=93\% and P$_{v+\pi}$=87\%), but as for HIP 12545 AB, there is a 
difference of only 3.1 km s$^{-1}$ between the predicted radial velocities 
for $\beta$PMG (10.3 km s$^{-1}$) and COL (7.2 km s$^{-1}$).
A membership in COL is favored since the Galactic position of this star is 
closer to that of COL compared to $\beta$PMG.

{\bf HIP 95261 AB}:
This binary A0V+M7V was originally proposed by \citet{2000zuckermanwebb} to be 
a member of THA, but its membership was revised to $\beta$PMG one
year later by \citet{2001bzuckerman}.
Without RV information, our analysis yields a membership probability (P$_{\pi}$)
of 99\% in $\beta$PMG but it decreases to 0\% when RV and parallax information 
are included (P$_{v+\pi}$). 
The reason for this drop is its radial velocity (13.0$\pm$2.5 km s$^{-1}$)
that is 15 km s$^{-1}$ off from that predicted for $\beta$PMG 
(-1.0$\pm$2.0 km s$^{-1}$).
Since this star has a low-mass (M7V) companion that could also affect 
the systemic velocity, the current RV measurement should be taken with
caution.
%, the effect is unlikely to be as high as 15 km s$^{-1}$. 
Another likely explanation is that the RV may be erroneous since 
this star is a very
fast rotator \citep[vsini = 330 km s$^{-1}$ ;][]{2009dasilva}.
To confirm this hypothesis, spectroscopic follow-up of 
the companion should be sought, in particular to search for low surface 
gravity indicators which should be significant if the star is a member of $\beta$PMG.
Also, new RV measurements would be highly desirable to 
better determine the systemic velocity of the system. 

{\bf 2MASSJ06085283-2753583} :
This young brown dwarf (M8.5) was proposed by \citet{2010rice} 
to be a member of $\beta$PMG which is confirmed by our 
analysis (P$_{v}$=75\% and P$_{v+\pi}$=95\%). 
We should note that our analysis is not optimal for this object 
because its $I_{c}-J$ > 3.25 is beyond the limit of our sequences. 
If we do not consider the photometry in our analysis, but 
consider the radial velocity and the trigonometric distance, the membership 
probability is 93\% in $\beta$PMG.
Since the IR spectrum shows telltale signatures of low-gravity 
($H$-band shape, KI lines), this target should be warranted the status 
of {\em bona fide member} of $\beta$PMG.

{\bf HIP 23418 ABCD} :
This quadruple system was proposed to 
be in $\beta$PMG by \citet{2003song} but does not appear in 
our {\em bona fide member} list because the uncertainty 
on its Hipparcos trigonometric distance is larger than our adopted criterion (5$\sigma$). 
Recently, a better parallax was measured by Riedel et al. (in prep) 
at 40.18$\pm$2.07\,mas.
This star has a membership probability of P$_{v}$=99\% to be member
of $\beta$PMG (see 2MASSJ05015881+0958587; Table~\ref{tab:candprop}) and
P$_{v+\pi}$=99\%. 
This system should be included in the list of {\em bona fide members} of 
$\beta$PMG.

{\bf HIP 50156}:
This star (M1V) was proposed by \citet{2012schlieder} to be 
a member of $\beta$PMG. 
From our analysis, this star seems to be a member of COL (P$_{v+\pi}$=93\%).
However, the radial velocity of this star is somewhat ambiguous: 6.9$\pm$1.0
\citep{2007kharchenko}, 10$\pm$1 \citep{2001montes}, 5.8$\pm$0.8 \citep{2006gontcharov} 
and 2.7$\pm$0.1 \citep{2010lopez} km s$^{-1}$.
It is unclear why the RVs differ so much from these measurements.
One possibility is that their radial velocity measurements are skewed by 
(unknown) binary orbital motion. 
The predicted radial velocity is 3.2, 8.4 and 1.3 km s$^{-1}$ for 
membership in $\beta$PMG, COL and ABDMG, respectively.
Spectroscopy from \citet{2009shkolnik} and \citet{2010lopez} did not unveil
any lithium detection which would point to a relatively old association
such as ABDMG. 
If we do not consider the radial velocity information in our analysis, but 
still consider the photometry and the trigonometric distance, the membership 
probability (P$_{\pi}$) is 99\% in COL. 
This star appears young but more RV measurements are desirable 
to better assess its membership.

\subsubsection{TWA}
{\bf TWA 19}:
This star (G5V) was rejected as a member of TWA and suggested instead to be
member of the Scorpius-Centaurus complex by \citet{2005mamajek} and 
\citet{2008torres}. 
Recently, \citet{2011chen} proposed this star as a member of 
Lower Centaurus Crux (LCC).
While our analysis yields a membership probability in TWA of
P=83\%, (P$_{v}$=94\%, P$_{v+\pi}$=75\%), it would be interesting to add the
Scorpius-Centaurus complex into our analysis but this is beyond the scope of 
the current paper since we restrict our search to relatively nearby co-moving groups 
within 100\,pc from the Sun. 

\subsubsection{THA}
{\bf HIP 14551} :
This star (A5V) was proposed by \citet{2011zuckerman} to be 
a member of THA. 
From our analysis, this star has an ambiguous status, with a 
membership probability P$_{v+\pi}$=50\% in THA and 49\% in COL.

{\bf HIP 17782 AB} :
This binary (G8V*) was proposed by \citet{2011zuckerman} to be 
a member of THA.
From our analysis, this star seems to be a member of COL (P$_{v+\pi}$=98\%).
The main reason for the higher probability in COL is the better agreement of
its Galactic positions $XZ$ with the members of COL. 
We propose that HIP 17782 AB be assigned as {\em bona fide member} 
of COL.

{\bf HIP 2484 AbB}: 
This star was proposed by \citet{2004zuckerman} to be a member of THA.
From our analysis, this star seems to be a member of THA (P$_{v+\pi}$=90\%), 
however the probability without parallax (P$_{v}$=48\%) is low 
because there is a 7 km s$^{-1}$ difference between the measured 
and predicted radial velocities. 
We should note that HIP 2484 AbB and HIP 2487 AB form a quintuple system. 
The measured and predicted
radial velocities for HIP 2487 AB are almost the same. 
We thus conclude that the measured
radial velocity of HIP 2484 AbB is affected by the binary components and that
HIP 2484 AbB is very likely a {\em bona fide member} of THA.

{\bf HIP 105404 AB} : 
This binary (G9V(eb)) was proposed by \citet{2004zuckerman} 
to be a member of THA.
Our analysis also assigns this star in THA (P=99\%) with a higher probability 
to be a binary, which is indeed the case. 
However, the membership (P$_{v}$) decreases to 0\% when the RV is included. 
The reason is that measured RV is incompatible with the association, 
differing by 8 km s$^{-1}$ from the predicted value.
However, the companion of this star could have an impact on the measured
radial velocity. 
If we do not consider
the radial velocity information in our analysis, but still consider 
the photometry and the
trigonometric distance, the membership probability is 99\% in THA. 
\citet{2008mentuch} measured an EW$_{\rm Li}$=171$\pm$22 m\AA\ suggestive of youth.
This star is also known to have a circumstellar disk detected at 24 $\micron$
\citep{2011zuckerman}.
More radial velocity monitoring of this star is needed to better assess its
membership. 

{\bf HIP 3556, HIP 9685, HIP 104308}:
These stars (M3V, F2V, A5V) were proposed by \citet{2004zuckerman} 
to be members of THA.
From our analysis, the inclusion of the radial velocity causes a 
decrease of their membership probabilities. 
For HIP 3556, HIP 9685 and HIP 104308, there is a 7-8 km s$^{-1}$ difference 
between the measured and predicted radial velocities. 
It is unclear why the RV differ so much from the expected values. 
One possibility is that their radial velocity measurements are skewed 
by (unknown) binary orbital motion, whereas the true system radial 
velocities would be consistent with the predicted values. 
However, no binary component are known to be orbiting around 
these stars. 
It is also possible that these objects just happen to have space motions 
relatively far from the bulk motion of the association. 
We should note that there is a rather large uncertainty on the radial velocity 
measurements for these stars. 
For HIP 104308, \citet{2000zuckermanwebb} noted that the measurement of the 
radial velocity was difficult (early type star, vsini $>$ 100 km s$^{-1}$).
To confirm the membership of these stars, we need better radial velocity 
monitoring. 

{\bf HIP 83494 and HIP 84642 AB} :
These stars (A5V, G8V+M5V) were proposed by \citet{2011zuckerman} to 
be members of THA. 
From our analysis, these stars seem to be field dwarfs. 
The main reason is that the Galactic positions ($XYZ$) of 
these stars are very different of the bulk position of the association. 
These stars are young as revealed by the presence of a circumstellar disk 
\citep{2011zuckerman} and lithium absorption for HIP 84642 AB 
\citep[228 m\AA;][]{2006torres}.
It is possible that
these objects just happen to have space motions relatively far
from the bulk motion of the associations considered in this work.

\subsubsection{COL}
{\bf HIP 12413 A} :
This star (A1V) was proposed by \citet{2011zuckerman} to be a member of COL. 
From our analysis, this star seems to be a member of $\beta$PMG (P$_{v+\pi}$=96\%).
There is a difference of only 0.9 km s$^{-1}$ between the predicted radial 
velocities for $\beta$PMG (16.6 km s$^{-1}$) and COL (15.7 km s$^{-1}$).
If we do not consider the RV information in our analysis, but still consider
the photometry and trigonometric distance, the membership probability 
is 88\% in THA.
\citet{2011zuckerman} note that HIP 12413 is a triple system 
(HIP 12413 A is a tight binary and HIP 12413 C a more distant 
($\rho$=25$^{\prime\prime}$) M dwarf). 
Spectroscopic follow-up of the M dwarf, in particular to measure the Li
absorption, would be very useful for constraining the age and the membership
of this system.

{\bf HIP 30030} :
This star (G0V) was proposed by \citet{2004zuckerman} to be member of THA. 
From our analysis, this star has P$_{v+\pi}$=85\% in COL.
There is a difference of only 2.9 km s$^{-1}$ between the predicted radial 
velocities for THA (19.3 km s$^{-1}$) and COL (22.2 km s$^{-1}$).
The main reason for the higher probability in COL is its Galactic positions 
XYZ which are much similar to COL than THA.

\subsubsection{CAR}
{\bf HIP 30034 A} :
This star (K1Ve) was originally proposed by \citet{2004zuckerman} to be a 
member of THA and by \citet{2008torres} to be a member of CAR. 
Our analysis favors a membership in CAR albeit with a modest probability 
(P$_{v+\pi}$=71\%, 27\% in COL). 
There is a difference of only 1.0 km s$^{-1}$ between the predicted radial 
velocities for COL (22.0 km s$^{-1}$) and CAR (23.0 km s$^{-1}$).
The main reason for the higher probability in CAR is the Galactic positions 
YZ that are more similar to CAR than THA.

\subsubsection{ARG}

{\bf HIP 4448 AB} :
This binary (K3Ve+K4Ve) was proposed by \citet{2008torres} to be member of ARG
which is confirmed by our analysis (P$_{v+\pi}$=86\%).
The probability below 90\% is explained by its $U$ velocity of
-16.1 km s$^{-1}$ which is somewhat different from the average value of
-22 km s$^{-1}$ for ARG {\em bona fide members}.

{\bf HIP 57632}:
This star (A3V) was proposed by \citet{2011zuckerman} to be a member of ARG 
which is confirmed by our analysis albeit with a modest probability 
(P$_{v+\pi}$=77\%; 22\% in the field). 
The relatively large uncertainty on the $J$ band photometry does not explain
this low probability since the latter remains low even when the photometry is
excluded from the analysis.
The main reason for its low probability is the Galactic space velocity $V$ 
of this star ($V$=-16.0) which has a difference 
of 4 km s$^{-1}$ with the average of ARG {\em bona fide members}. 

\subsubsection{ABDMG}

{\bf HIP 93580}:
This star (A4V) was proposed by \citet{2011zuckerman} to be a member of ABDMG.
Our analysis also puts this object in ABDMG, but with a somewhat low probability
(P$_{v+\pi}$=80\%).
The Galactic space velocity $U$ (-11.3 km s$^{-1}$) is somewhat 
different from the bulk motion of ABMDG members ($U$=-7.1 km s$^{-1}$). 
It is possible that this object just 
happens to have space motions relatively far from 
the bulk motion of the association. 

{\bf HIP 117452 AB}:
This binary (A0V*) was proposed by \citet{2011zuckerman} to be a member of ABDMG.
From our analysis, this star has a membership probability of P$_{v+\pi}$=92\% 
in the field.
We should note that our analysis is not optimal for this object 
because its $I_{c}-J$ < -0.1 is beyond the limit of our calculated sequences.
If we do not consider the photometry in our analysis, but 
consider the radial velocity and the trigonometric distance, the membership 
probability changes to 99\% in ABDMG.
\citet{2011zuckerman} detected a circumstellar disk excess 
at 24 and 70 $\micron$.
The tertiary companon of this system, HD 223340, is an early-K-type star 
about 75$\arcsec$ away that
shows Li absorption with an EW$_{\rm Li}$=148 m\AA\,\citep{2011zuckerman} 
that is intermediate between that observed in stars of similar spectral types
in ABDMG and THA \citep[see Fig.5; ][]{2008mentuch}.
Thus, we confirm HIP 117452 AB as a {\em bona fide member} of ABMDG. 

{\bf HIP 14807}:
This star was proposed by \citet{2004zuckerman} to be a member of ABDMG.
HIP 14809 and HIP 14807 are a binary system. 
HIP 14807 is not listed as a
{\em bona fide member} because no $I_{c}$ and RV measurements are availables 
for this star.
However, if we do not consider the RV and photometry information in our 
analysis, but still consider the trigonometric distance, the membership 
probability is 96\% in ABDMG.
Furthermore, the other binary component (HIP 14809) is confirmed to be in
ABDMG with P$_{v+\pi}$=99\%.
Our analysis confirms that both stars are {\em bona fide members} of ABDMG.

\subsection{Candidate members}
To confirm the membership of our 164 candidate members, we should include 
other observables in our analysis, such as radial velocity or parallax 
measurements.
In addition to performing our own radial velocity measurements for a 
number of our candidates, we searched in the literature and compiled all 
radial velocity measurements from the previous
studies of \citet{1996hawley,2001montes,2002gizis,2006torres, 2006moor, 
2009lepine, 2010schlieder, 2010looper, 2011kiss, 2011rodriguez, 2011rave, 
2012bowler, 2012shkolnik, 2012schlieder, 2012bschlieder}.
A good RV measurement ($\pm$2 km s$^{-1}$, not known to be spectroscopic
binaries) is available for 35 of our 164 candidates, and 
Table~\ref{tab:allcand} shows the membership probability before and after 
the inclusion of the radial velocity in our Bayesian analysis.
In general, the membership probability increases or remains high after 
inclusion of the radial 
velocity measurement; thus, these stars (21/35) are strong 
candidate members of these seven young kinematic groups. 
For the other stars, the radial velocity measurement differs from those 
predicted for the seven young kinematic groups and need spectroscopic 
follow-up to search for various youth indicators. 
We searched the literature and compiled all available parallax 
measurements; we found this information for 16 candidates 
either from the literature \citep{2007vanleeuwen, 2011wahhaj, 2012shkolnik} 
and ongoing work (Riedel et al.), and Table~\ref{tab:allcand} shows the membership probability
before and after the inclusion of the radial velocity and the parallax in our Bayesian analysis..
Finally, we also compiled all available Li measurements, 
in addition to our own measurements; Li was measured for 12 candidates. 
Here we discuss the most promising candidates deserving more observations to
confirm their status as young stars.
We should note that full
confirmation of the membership of these candidate members will require 
complete $UVWXYZ$, hence accurate proper motion, radial velocity 
and parallax measurements and observation of signs of youth.

\subsubsection{New highly probable candidate members}

Of our 164 candidate members with P $>$ 90\%, 35 have a
radial velocity measurement (of which 14 are from our work), 12 have Li
detection (2 from our work), and 16 have a trigonometric parallax measurement.
All new parallax measurements are compiled in Table~\ref{tab:candprop}. 
In general, there is a good agreement between the statistical distance inferred from our analysis 
and the true trigonometric one (see Table~\ref{tab:plxcand}). 
We discuss below those candidate members with P$_{v}$ $>$ 90\% and the presence of
sign of youth or P$_{\pi}$ $<$ 90\% or P$_{v+\pi}$ $>$ 90\% .

{\bf 2MASSJ01112542+1526214 AB (GJ 3076)} : 
We propose this binary 
\citep[$\rho$=0.41$^{\prime\prime}$, M5V + M6V;][]{2004beuzit,2012janson} as a 
highly probable member of $\beta$PMG with a P$_{v}$=99.99\% at d$_{s}$=20$\pm1$\,pc.
Also, our analysis predicted the binary status of this star.
Recently, a trigonometric distance (21.8$\pm$0.8\,pc) was measured by 
Riedel et al. (in prep), in very good agreement with our predicted statistical distance.
Including this parallax in our analysis yields a P$_{v+\pi}$=99.99\% (binary) for
$\beta$PMG (see Table~\ref{tab:plxcand}). 
Our ESPaDOnS spectrum of 2MASSJ01112542+1526214 AB (unresolved) 
is shown in figure~\ref{fig:lithium}.
It shows a strong EW$_{\rm Li}$ of 629$\pm$14\,m\AA\, which  
is quite high for a candidate member of $\beta$PMG 
although there is no other known late-type dwarf in 
$\beta$PMG for comparison. 
Note that \citet{2001montes} proposed this star as a young disk star, but did not
assign it to any particular association.
This star fits all requirements to be considered as a new 
{\em bona fide member} of $\beta$PMG.

{\bf 2MASSJ05241914-1601153 AB} :
We propose this binary as a highly probable member of $\beta$PMG with a 
P$_{v}$=99.9\% at d$_{s}$=20$\pm5$\,pc.
This candidate is a resolved binary system (M4.5V + M5.0V) that was discovered by 
\citealt{2010bergfors}, with a separation of 0.639$\pm$0.001$^{\prime\prime}$.
Our ESPaDOnS spectrum of 2MASSJ05241914-1601153 (unresolved) is shown 
in Figure~\ref{fig:lithium}.
Li is clearly detected with an equivalent width of 223$\pm$27 m\AA.
The broadened absorption line is an effect of the fast rotation of the star.
This system is similar to the {\em bona fide $\beta$PMG member} M4Ve + M4.5V 
binary system HIP 112312 AB for which the latest spectral type object shows 
lithium absorption with an EW$_{\rm Li}$=315$\pm$22 m\AA\, \citep{2008mentuch}.
This is a very strong candidate member of $\beta$PMG, and only a trigonometric distance 
measurement is needed to fully confirm its {\em bona fide} status.  

{\bf 2MASSJ05332558-5117131 (TYC 8098-414-1)} : 
We proposed this K7Ve star as a highly probable candidate member of THA with 
P$_{v}$=99.9\% at d$_{s}$=54$\pm4$\,pc.
This candidate seems to be a young dwarf because \citet{2006torres}
measured an EW$_{\rm Li}$= 50 m\AA\, consistent with other {\em bona fide members} 
of THA for an age between 10 and 40\,Myr.
A trigonometric parallax measurement is needed to confirm the {\em bona fide}
 status of this star.

{\bf 2MASSJ05531299-4505119}:
We propose this M0.5V star as a highly probable candidate member of ABDMG with 
P$_{v}$=99.9\% at d$_{s}$=34$\pm4$\,pc.
This candidate seems to be a young dwarf because \citet{2006torres}
measured an EW$_{\rm Li}$= 140 m\AA\, which  
is quite high for a candidate member of ABDMG.
A trigonometric distance 
measurement and other youth indicators 
are needed to confirm its {\em bona fide} status.

{\bf 2MASSJ18202275-1011131 (HIP 89874 AB)} :
From our analysis, this binary star \citep[K5Ve+K7Ve;][]{2010tetzlaff} 
seems to be a highly probable 
candidate member of ARG with P=99\% at d$_{s}$=34$\pm$4\,pc.
The membership probability with RV information is doubtful because there 
is a difference of 4.1 km s$^{-1}$ between three published radial 
velocity measurements \citep[-13.8, -9.7, -9.0 km s$^{-1}$][]{2006torres,2010lopez,2001montes}.
This large difference may come from the binary, or else it may reflect 
a large uncertainty due to 
the relatively fast rotation of the star 
\citep[vsini = 20.1 km s$^{-1}$;][]{2010lopez}.
If we repeat our analysis with RV information (-13.8 km s$^{-1}$) and
parallax measurement \citep[13.17$\pm$3.81 mas;][]{2007vanleeuwen}, the 
membership probability changes to 94\% (binary) in $\beta$PMG.
The predicted radial velocity is -24.8 km s$^{-1}$ for ARG and 
-16.2 km s$^{-1}$ for $\beta$PMG.
Also, this star has a relatively strong EW$_{\rm Li}$ of 530 m\AA\ 
\citep{2006torres} suggestive an age of 5-15\,Myr. 
\citet{2001montes} proposed this star as a young disk member. 
The Galactic positions (71.2, 26.4, 2.9\,pc) of this star is very different
from those of {\em bona fide members} of young co-moving groups listed in
Table~\ref{tab:membprop}. 
Could this star be a member of the Scorpius-Centaurus complex?
\citet{2001mamajek} suggest that this is not the case since the star 
never had a close passage near Sco-Cen in the past.

{\bf 2MASSJ13591045-1950034 (GJ 3820) }:
Without RV information, our analysis assigns this M4.5V flare star 
\citep{2002gizis} in ARG with P=95.6\% at a statistical distance 
of 8$\pm1$\,pc, in fair agreement with the measured parallax of 
10.7$\pm$0.1\,pc (Riedel, in prep).
This star also has a measured RV of -15.8 km s$^{-1}$ \citep{2002gizis}.
Including this RV and parallax in our analysis yields P$_{v}$=85\% and 
P$_{v+\pi}$=99\%, both in the field.
This membership probability should be taken with caution, since 
\citet{1996hawley} measured a RV of 35$\pm$10 km s$^{-1}$ for this star, 
which is very different from the RV measurement of \citet{2002gizis}.
These measurements suggest that this star may be a spectroscopic binary 
(sb1).
Thus, although we cannot exclude that this star is young, it does not appear
to be a likely member of any of the co-moving groups considered in
this work.
More RV monitoring is required to determine if this system is a binary.
A measurement of EW$_{\rm Li}$ would also provide a very useful 
constraint on the age.

{\bf 2MASSJ00503319+2449009 (HIP 3937), 2MASSJ03033668-2535329 (HIP 14239)} :
Our analysis suggests membership in ARG for both stars with P=99.9\%.
However, each have a trigonometric distance in disagreement with our 
statistical distances (see Tables~\ref{tab:candall},~\ref{tab:allcand}).
When the parallax and radial velocity information are included into the analysis, 
both stars appear in the field (J0303-2535, P$_{v+\pi}$=99\%; J0050+2449, P$_{v+\pi}$=99\%).

{\bf 2MASSJ23301341-2023271 (HIP 116003 AB) } :
Our analysis suggests this M3V* star to be a candidate 
member of COL with P$_{v}$=76\%. 
The P$_{v}$ should be taken with caution as this star is a known spectroscopic
binary \citep{2006torres} which displays important RV variations. 
Indeed, \citet{2002gizis} measured four times the RV of this star 
(11.7, -4.7, -16.7 and 30.1 km s$^{-1}$), and 
\citet{2006torres} measured it three times with a mean RV=-5.7 km s$^{-1}$.
The range of measured RVs includes our predicted value (-2.9 km s$^{-1}$) for 
COL.
When the trigonometric distance \citep[61.72$\pm$3.53 mas;][]{2007vanleeuwen}
is taken into account, but excluding the RV information for the reason
just mentioned, P$_{\pi}$=99\% (binary) in $\beta$PMG. 
\citet{2006torres} did not detect lithium absorption, however the quality
of their spectrum is not good. 
Better measurements of age-dating indicators and additional radial velocity 
monitoring of this star would be useful to verify its status. 

{\bf 2MASSJ10121768-0344441 (HIP 49986)}:
Our analysis suggests this (M1.5V) to be candidate member of ABDMG with
P$_{v}$=91.4\%.
A parallax measurement is available for this star and the measured 
distance differs significatly from our predicted value (see Table~\ref{tab:plxcand}).
Thus, this star appears to be in the field with P$_{v+\pi}$=99\%.
In addition, no youth indicators have been measured for this star yet.

{\bf 2MASSJ01351393-0712517, 2MASSJ01365516-0647379, 
2MASSJ05254166-0909123, 2MASSJ20434114-2433534 and 2MASSJ23205766-0147373} :
Our analysis suggests these stars to be candidate members of 
$\beta$PMG (J0135-0712,M4V; J0136-0647,M4V; J2043-2433,M3.7+M4.1),
ARG (J2320-0147,M4+M4) and ABDMG (J0525-0909,M3.5+M4) with a membership probability of 
P$_{v+\pi} >$ 99\% (see Table~\ref{tab:plxcand}). 
Only age-dating indicators are needed to fully confirm their 
{\em bona fide} status. 

{\bf 2MASSJ06131330-2742054, 2MASSJ20100002-2801410 and 2MASSJ20333759-2556521}:
From our analysis, we proposed these (M3.5; M2.5+M3.5; M4.5) stars to be candidate members of 
$\beta$PMG with a membership probability of P=99\%. 
No RV measurement is available for these stars, however a trigonometric 
distance was measured by Riedel (in prep) for each star.
When the parallax information is included into the analysis, they all
appear to be in $\beta$PMG with P$_{\pi} >$ 99\%.
To confirm the membership of these stars, we need 
measurements of radial velocity and age-dating indicators. 

%%%%%%%%%%%%%%%%%%%%%%%%%%%%%%%%%%%%%%%%%%%%%%%%%%%%%%%%%%%%%%%%%%%%%%%%%%%%%%%%%%%%%%%%%%%%%%%%%%%%%%%%%%%%%
\subsubsection{Previously identified in the literature}
%%%%%%%%%%%%%%%%%%%%%%%%%%%%%%%%%%%%%%%%%%%%%%%%%%%%%%%%%%%%%%%%%%%%%%%%%%%%%%%%%%%%%%%%%%%%%%%%%%%%%%%%%%%%%

Of the 71 candidate members of the seven young kinematic groups identified 
in the literature, 58 have P $>$ 90\% or P$_{v}$ $>$ 90\% 
(see table~\ref{tab:allcand}); those are thus likely to be true members of the
associations. 
We discuss below the remaining 13 stars and also those star for which we 
assign an association membership different from that previously proposed in 
the literature.
Also, we discussed 5 new {\em bona fide members} of $\beta$PMG (2) and ABDMG (3). 

{\bf 2MASSJ00233468+2014282 (TYC 1186-706-1)}:
This star (K7.5V) was proposed by \citet{2009lepine} to be a member 
of $\beta$PMG.
From our analysis, this star has an ambiguous membership (P$_{v}$) of 55\% 
in the field and 35\% in COL.
If we exclude the photometry of the star from our analysis, but include 
the RV information, then it has a membership probability of 85\% in the field. 
To confirm the membership of this star, we need 
a trigonometric distance and measurements of age-dating indicators. 

{\bf 2MASSJ01220441-3337036 (TYC 7002-2219-1) }:
This star (K7Ve) was proposed by \citet{2010schlieder} to be candidate member 
of ABDMG. 
From our analysis, this star is a candidate member of THA (P$_{v}$ =99\%) due
to RVs \citep[4.8, 3.0$\pm$1.4 km s$^{-1}$;][]{2006torres,2010schlieder} 
much closer to that predicted for THA (4.5$\pm$1.3 km s$^{-1}$) compared 
to ABDMG (18.2$\pm$2.1 km s$^{-1}$). 
\citet{2006torres} did not detect lithium absorption for this star, 
however there is no other K7V in THA for comparison. 
To confirm our hypothesis, we need a trigonometric distance measurement.

{\bf 2MASSJ03241504-5901125} :
\citet{2000torres} and \citet{2004delareza} proposed this star (K7V) as a 
candidate member of THA.
From our analysis, this star seems to be a highly probable candidate 
member of COL (P$_{v}$=99\%) due to a RV 
\citep[17.5$\pm$1.3 km s$^{-1}$;][]{2006torres} much closer
to that predicted for COL (17.6$\pm$1.0 km s$^{-1}$ ) compared to THA 
(13.8$\pm$1.6 km s$^{-1}$).
This candidate seems to be a young star with an EW$_{\rm Li}$ = 235 m\AA\ 
\citep{2006torres}, consistent with the {\em bona fide members} of 
young associations with an age less than 40\,Myr.

{\bf 2MASSJ09361593+3731456 (HIP 47133)}:
This binary star (M2+M2) was proposed by \citet{2012schlieder} to be a member of the 
$\beta$PMG.
From our analysis, this star is a likely member of the field with
P=99\%, P$_{v}$=99\%, P$_{\pi}$=99\% and P$_{v+\pi}$=99\%.
\citep{2012bschlieder} found that this star is a spectroscopic binary (M2+M2) and 
have measured a RV of -2.5$\pm$1.0 km s$^{-1}$).
The main reason for this star to be in the field is that its Galactic 
position $Z$ (25.1\,pc) is far (39\,pc away) from the center of 
$\beta$PMG {\em bona fide members}.
A measurement of EW$_{\rm Li}$ would also provide a very useful 
constraint on the age.

{\bf 2MASSJ11254754-4410267 }:
This binary M4+M4.5 star was proposed by \citet{2011rodriguez} to be a candidate
member of TWA.
From our analysis, this star has a membership probability of 
P$_{v}$=99.9\% (binary) in ABDMG.
This star has an EW$_{\rm Li}$ = $<$30 m\AA\ \citep{2011rodriguez}, which
is consistent with members of ABDMG of similar spectral type.
A parallax measurement is needed to confirm the membership of this star in ABDMG.

{\bf 2MASSJ11455177-5520456 (CD-54 4320)}:
This star (K5Ve) was proposed by \citet{2008torres} to be a member of CAR. 
From our analysis, this star has an ambiguous membership (P$_{v}$) of 46\% 
in TWA and 48\% in COL.
The measured RV (16.1 km s$^{-1}$) falls slightly closer to the predicted
value for COL, although there is a difference of only 2.0 km s$^{-1}$ 
between the predicted radial 
velocities for TWA (11.8 km s$^{-1}$) and COL (13.9 km s$^{-1}$).
\citet{2009dasilva} measured an EW$_{\rm Li}$=190\,m\AA, which confirm
its age similar to THA/COL/CAR {\em bona fide members}.

{\bf 2MASSJ11493184-7851011 (V* DZ Cha)} :
\citet{2008torres} proposed this star (M1V) as a candidate member of 
$\epsilon$ Cha. 
From our analysis, this star is a highly probable candidate member 
of $\beta$PMG with P$_{v}$=91\%.
Our analysis gives a statistical distance of 71$\pm$6\,pc, which would make this star 
as one of the most distant members of $\beta$PMG. 
As noted earlier, caution should be applied for our candidates with large statistical distances 
as we did not include the associations beyond 100\,pc in our analysis. 
It is thus possible that this star is a member of the more distant ($\sim$110\,pc) 
$\epsilon$ Cha association ($\sim$8 Myr), whose spatial velocity is close to that of $\beta$PMG. 
The EW$_{\rm Li}$ of this star was measured by \citet[][;560 m\AA]{2009dasilva}, 
and it is somewhat larger than those of $\beta$PMG members of similar spectral types.  
This is clearly a young star, but a parallax measurement is needed to confirm its membership. 
And ideally we should add the $\epsilon$ Cha association in our analysis. 

{\bf 2MASSJ12151838-0237283 (TYC 4943-192-1)}:
This star (M0Ve) was proposed by \citet{2010schlieder} to be a member of ABDMG.
From our analysis, this star has an ambiguous membership (P$_{v}$) of 83\% in 
ABDMG and 17\% in the field.
Although a promising candidate for ABDMG, age-dating
indicators and a trigonometric parallax are needed to confirm its membership. 

{\bf TWA 15 A (2MASSJ12342064-4815135) and TWA 15 B (2MASSJ12342047-4815195)}
These stars (M1.5V and M2V) were proposed by \citet{2006delareza} 
to be members of TWA. 
\citet{2008torres} rejected these stars as TWA members but suggested that they
could be members of the Scorpius-Centaurus complex.
From our analysis, these stars have 
membership probabilities (P) of 99\% and P$_{v}$ =99\% in the field.
We repeated our analysis without considering the photometry of the star and 
the membership probabilities are still 99\% for the field. 
However, these stars appear to be young since \citet{2008mentuch} 
unveiled lithium absorption for all of them  
(EW$_{\rm Li}$=494 and 484 m\AA\, for TWA 15A and 15B, respectively).
To confirm their membership, we should add in our analysis the 
Scorpius-Centaurus complex. 

{\bf TWA 18 (2MASSJ13213722-4421518)}:
This star (M0Ve) was proposed by \citet{2006delareza} to be a member of TWA.
From our analysis, this candidate member has a membership 
probability (P$_{v}$) of 89.2\% at d$_{s}$=60$\pm$5\,pc for TWA. 
This star is confirmed to be young since \citet{2008mentuch} measured
an EW$_{\rm Li}$= 464\,m\AA, consistent with the age of TWA. 
A trigonometric distance measurement is needed to confirm its membership. 

{\bf 2MASSJ16430128-1754274}:
This star (M0.5V) was proposed by \citet{2011kiss} to be a member of 
$\beta$PMG.
From our analysis, this star has an ambiguous membership (P$_{v}$) of 65\% in 
the field and 35\% in $\beta$PMG.
However, \citet{2011kiss} measured EW$_{\rm Li}$=300$\pm$20 m\AA\, confirming 
that the age of this star is similar to that of $\beta$PMG 
{\em bona fide members}.
To confirm the membership, we need a trigonometric distance measurement.

{\bf 2MASSJ20072376-5147272 (CD-52 9381)}:
This star (K6Ve) was proposed by \citet{2008torres} to be a member of ARG.
From our analysis, this star has a membership probability (P$_{v}$) 
of 89.5\% in ARG at d$_{s}$=31$\pm$1\,pc.
This star shows EW$_{\rm Li}$=60 m\AA\, \citep{2009dasilva}, consistent with
an age less than 70 Myr.
To confirm the membership of this star, we need a parallax measurement.

{\bf 2MASSJ21212446-6654573 (HIP 105441 A,B)}:
This system (K2V+K7V, $\rho$=26$^{\prime\prime}$) was proposed by 
\citet{2001tzuckerman} to be a member of THA, and by \citet{2009ortega} 
to be a member of $\beta$PMG.
From our analysis, this system seems to be a member of $\beta$PMG 
with P$_{v}$=99\%. 
If we take into account the trigonometric distance of HIP 105441 A 
\citep[$\pi$=33.14$\pm$1.45 mas;][]{2007vanleeuwen},
the membership probability P$_{v+\pi}$ is 99\% for $\beta$PMG.
We should note that its radial velocity is quite uncertain 
from previous studies (A:6.4$\pm$14.8,-24.1,\citet{2007kharchenko, 2006torres}, 
B:3.3 km s$^{-1}$,\citet{2006torres}.
HIP 105441 A could thus be a spectroscopic binary \citep{2006torres}.
The predicted radial velocity measurement is  
5.7 km s$^{-1}$ for $\beta$PMG.
When the trigonometric distance is taken into account, but excluding the 
RV information for the reason just mentioned, P$_{\pi}$=99\% in $\beta$PMG. 
However, \citet{2006torres} showed that HIP 105441 A does not have a
Li detection but measured an EW$_{\rm Li}$=15 m\AA\ for 
HIP 105441 B (K7V). 
To confirm membership, we need more radial velocity measurements and 
other age-dating measurements for the two stars.

{\bf 2MASSJ22424884+1330532 (TYC 1158-1185-1 N,S)}:
This binary (K5Ve*) was proposed by \citet{2010schlieder} to be a candidate 
member of $\beta$PMG, however this star was ruled out after RV measurement.
From our analysis, this star has an ambiguous membership (P) of 
46\% in $\beta$PMG and 48\% in the field.
There is a difference of only 0.7 km s$^{-1}$ between the predicted radial 
velocities for $\beta$PMG (-8.0 km s$^{-1}$) and the field (-7.3 km s$^{-1}$).
If we exclude the photometry of the star from our analysis, but include 
the RV information, then it has a membership probability of 60\% in COL and
40\% in the field.
The RV measurement (-14.9 km s$^{-1}$) is much closer to the predicted 
radial velocity for COL (-14.4 km s$^{-1}$) compared to the field 
or $\beta$PMG.
To confirm the membership, we need a trigonometric distance and 
measurements of age-dating indicators.

{\bf 2MASSJ05064991-2135091 A, 2MASSJ05064946-2135038 BC (GJ 3332)} :
This close visual binary 
\citep[A: M1Ve and BC:M3.5V+M4V, $\rho$=1.2$^{\prime\prime}$;][]{2008torres} 
was proposed by \citet{2008torres} to be a member of $\beta$PMG.
From our analysis, the membership probability P$_{v}$ is 96\% in COL for
the component A and 99\% in $\beta$PMG (component BC). 
The higher probability in COL for component A is mainly due to a difference
in space velocity, which could be the result of orbital motion.
This system is known to be young as \citet{2009dasilva} 
measured EW$_{\rm Li}$=20 m\AA\,, for both components A and BC; these values are 
consistent with the age of $\beta$PMG, and the component's spectral types, 
although it is a bit low for the A component.
Recently, a trigonometric distance (19.2$\pm$0.5\,pc) was measured 
by Riedel (in prep) which is consistent with our predicted statistical distances 
for $\beta$PMG.
With the parallax measurement included into our analysis, both components
appear to be in $\beta$PMG with a very high membership probability of 
P$_{v+\pi}$=99.9\% (see Table~\ref{tab:plxcand}). 
This close visual binary fits all requirements to be considered as a new 
{\em bona fide member} of $\beta$PMG.

{\bf 2MASSJ06091922-3549311 (CD-35 2722)} :
We confirm the membership of this M0.5Ve star in ABDMG, as previously 
proposed by \citet{2008torres}.
Recently, the NICI Planet-Finding Campaign confirmed that this star has 
a L4$\pm1$ companion physically associated with the primary star
\citep{2011wahhaj}. 
From this study, a trigonometric distance was measured at 21.3$\pm$1.4\,pc.
From our analysis, we obtain P$_{v}$ = 99.9\% and P$_{v+\pi}$ = 99.9\% for ABDMG.
This star shows many signs of youth such as EW$_{\rm Li}$ =10 m\AA\,, X-ray and
H$\alpha$ emission, consistent with {\em bona fide members} of ABDMG.
This star fits all requirements to be considered as a new 
{\em bona fide member} of ABDMG. Thereby, the L4$\pm1$ companion is the
lowest mass {\em bona fide member} of ABDMG.

{\bf 2MASSJ04522441-1649219 (TYC 5899-26-1)}:
We confirm the membership of this M3Ve* star in ABDMG, as previously 
proposed by \citet{2008torres} and confirmed by \citet{2010schlieder} with
RV measurement.
Recently, a trigonometric distance was measured at 16.3$\pm$0.4\,pc by 
\citet{2012shkolnik}.
From our analysis, we obtain P$_{v}$ = 99.9\% (binary) and P$_{v+\pi}$ = 99.9\% 
(binary) for ABDMG.
This star shows many signs of youth such as EW$_{\rm Li}$ =20 m\AA\,, X-ray and
H$\alpha$ emission, consistent with {\em bona fide members} of ABDMG.
This star fits all requirements to be considered as a new 
{\em bona fide member} of ABDMG.

{\bf 2MASSJ21521039+0537356 (HIP 107948)}:
This star (M2Ve) was proposed by \citet{2008torres} to be a member of ABDMG.
\citet{2012shkolnik} measured a radial velocity of -15.1$\pm$1.5 km s$^{-1}$ 
and a trigonometric distance at 30.5$\pm$5.3\,pc.
With this new measurements, our analysis gives a P$_{v}$ = 99.9\% (binary) and 
P$_{v+\pi}$ = 99.9\% (binary) for ABDMG.
This star shows many signs of youth such as EW$_{\rm Li}$ =10 m\AA\,, X-ray and
H$\alpha$ emission, consistent with {\em bona fide members} of ABDMG.
\citet{2009shkolnik} estimated the age of this star between 20 and 
150\,Myr from various spectroscopic indices.
This star fits all requirements to be considered as a new 
{\em bona fide member} of ABDMG.

%%%%%%%%%%%%%%%%%%%%%%%%%%%%%%%%%%%%%%%%%%%%%%%%%%%%%%%%%%%%%%%%%%%%%%%%%%%%%%%%%%%%%%%%%%%%%%%%%%%%%%%%%%%%%%%%%%%%%%%
\section{SUMMARY AND CONCLUSION}
%%%%%%%%%%%%%%%%%%%%%%%%%%%%%%%%%%%%%%%%%%%%%%%%%%%%%%%%%%%%%%%%%%%%%%%%%%%%%%%%%%%%%%%%%%%%%%%%%%%%%%%%%%%%%%%%%%%%%%%
The study presented in this paper aims at extending 
the census of low-mass star members of seven 
young associations of the solar neighborhood: 
the $\beta$ Pictoris and AB Doradus moving groups, the TW Hydrae,  
Tucana-Horologium, Columba, Carina and Argus associations. 
To identifiy new members in these associations, we developed a Bayesian 
statistical analysis which, based on the values of 6 observables 
($I_{c}$ and $J$ magnitudes, amplitude of proper motion in right ascension 
and declination, right ascension and declination), computes a membership 
probability to a given association as well as the most probable 
(statistical) distance.

Starting from a sample of 758 stars, all showing indicators of youth
such as H$\alpha$ and X-rays emission, our analysis selected 164
young stars of spectral type between K5V and M5V with a membership probability 
over 90\%. We find one of theses candidates in the TWA, 37 in the $\beta$ PMG, 17 in the THA,
20 in the COL, 6 in the CAR, 50 in the ARG and 33 in the ABDMG. 
We also quantified the reliablity of our analysis by 
determining the recovery rate of known members of these associations;
the recevery rate is better than 90\%.
We also performed a Monte Carlo analysis to determine the expected 
number of false detections.
The false alarm rate is typically between 5 and 15\% depending on the
association, thus showing that our method is very effective at
identifying genuine members of young associations.

The kinematic models used in our statistical analysis predict the radial 
velocity of a candidate member, thus radial velocity follow-ups provide a powerful 
tool to further confirm membership. 
We initiated a program to measure the radial velocity of our new candidates.
The first results for 14 candidates show that adding their radial velocity
 measurement to the analysis yields probabilities above 99\%.
To clearly establish the young age of the candidate members of $\beta$PMG and THA, 
we initiated a program to detect the presence of photospheric lithium. 
Early results allowed us to detect lithium for 2 of 
our candidate members of $\beta$PMG: 2MASSJ05241914-1601153 (M4.5V+M5V) 
and 2MASSJ01112542+1526214 (M5V+M6V).
Also another of our candidates in THA, 2MASSJ05332558-5117131 (K7Ve), has both a
radial velocity confirmation (from \citealt{2006torres} and our work) and 
a lithium detection (from \citealt{2006torres}). 
Finally, we propose that six stars should be considered as new {\em bona fide} members
of $\beta$PMG (2MASSJ01112542+1526214, J05064991-2135091 and J05064946-2135038) 
and ABDMG (2MASSJ06091922-3549311, J04522441-1649219 and J21521039+0537356). 
%We have recently initiated a parallax program to  
%determine the trigonometric distance of our new candidates to better assess
%the membership of these stars.  

%%%%%%%%%%%%%%%%%%%%%%%%%%%%%%%%%%%%%%%%%%%%%%%%%%%%%%%%%%%%%%%%%%%%%%%%%%%%%%%%%%%%%%%%%%%%%%%%%%%%%%%%%%%%%%%%%%%%%%%
\acknowledgments
%%%%%%%%%%%%%%%%%%%%%%%%%%%%%%%%%%%%%%%%%%%%%%%%%%%%%%%%%%%%%%%%%%%%%%%%%%%%%%%%%%%%%%%%%%%%%%%%%%%%%%%%%%%%%%%%%%%%%%%

The authors would like to thank Emilie Storer for helping with 
the compilation of {\em bona fide} member lists, 
the night assistant J. Vasquez at the CTIO 0.9m, the
Gemini and CFHT staff for carrying out the observations, and 
Marie-Eve Naud for carefully reviewing the candidate member lists. 
Finally, we would like to thank our referee, Eric Mamajek, for excellent
suggestions that greatly improved the quality of this paper.

This work was supported in part through grants from the 
the Fond de Recherche Qu\'eb\'ecois - Nature et Technologie and
the Natural Science and Engineering Research Council of Canada. 
This research has made use of the SIMBAD database, operated at 
Centre de Donn\'ees astronomiques de Strasbourg (CDS), 
Strasbourg, France. 
This research has made use of the VizieR catalogue access 
tool, CDS, Strasbourg, France \citep{2000ochsenbein}.

Based on observations obtained at the Gemini Observatory, which is operated by the
Association of Universities for Research in Astronomy, Inc., under a cooperative agreement
with the NSF on behalf of the Gemini partnership: the National Science Foundation (United
States), the Science and Technology Facilities Council (United Kingdom), the
National Research Council (Canada), CONICYT (Chile), the Australian Research Council
(Australia), Minist\'{e}rio da Ci\^{e}ncia, Tecnologia e Inova\c{c}\~{a}o (Brazil) 
and Ministerio de Ciencia, Tecnolog\'{i}a e Innovaci\'{o}n Productiva (Argentina).

The DENIS project has been partly funded by the SCIENCE and the HCM plans of
the European Commission under grants CT920791 and CT940627.
It is supported by INSU, MEN and CNRS in France, by the State of Baden-W\"urttemberg 
in Germany, by DGICYT in Spain, by CNR in Italy, by FFwFBWF in Austria, by FAPESP in Brazil,
by OTKA grants F-4239 and F-013990 in Hungary, and by the ESO C\&EE grant A-04-046.
Jean Claude Renault from IAP was the Project manager.  Observations were  
carried out thanks to the contribution of numerous students and young 
scientists from all involved institutes, under the supervision of  P. Fouqu\'e,  
survey astronomer resident in Chile.  

Funding for RAVE has been provided by: the Australian Astronomical Observatory; 
the Leibniz-Institut fuer Astrophysik Potsdam (AIP); the Australian National University; 
the Australian Research Council; the French National Research Agency; the German 
Research Foundation (SPP 1177 and SFB 881); the European Research Council (ERC-StG 240271 Galactica); 
the Istituto Nazionale di Astrofisica at Padova; The Johns Hopkins University; the National 
Science Foundation of the USA (AST-0908326); the W. M. Keck foundation; the Macquarie University; 
the Netherlands Research School for Astronomy; the Natural Sciences and Engineering Research 
Council of Canada; the Slovenian Research Agency; the Swiss National Science Foundation; 
the Science \& Technology Facilities Council of the UK; Opticon; Strasbourg Observatory; 
and the Universities of Groningen, Heidelberg and Sydney. The RAVE web site is at 
\url{http://www.rave-survey.org}. 

%%%%%%%%%%%%%%%%%%%%%%%%%%%%%%%%%%%%%%%%%%%%%%%%%%%%%%%%%%%%%%%%%%%%%%%%%%%%%%%%%%%%%%%%%%%%%%%%%%%%%%%%%%%%%%%%%%%%%%%
\bibliographystyle{apj}
\bibliography{references2}
%%%%%%%%%%%%%%%%%%%%%%%%%%%%%%%%%%%%%%%%%%%%%%%%%%%%%%%%%%%%%%%%%%%%%%%%%%%%%%%%%%%%%%%%%%%%%%%%%%%%%%%%%%%%%%%%%%%%%%%

\clearpage
\clearpage
\onecolumngrid
\LongTables
\begin{landscape}
\tabletypesize{\tiny}
\begin{deluxetable}{llrrrrrrrrrrrrrrr}
\tablewidth{640pt}
\tablecolumns{17}
\tablecaption{{\em BONA FIDE MEMBERS} of young kinematic groups \label{tab:membprop}}
\tablehead{
\colhead{ } & \colhead{ } & \colhead{$\alpha$ (J2000.0)} & \colhead{$\delta$ (J2000.0)} & \colhead{$\mu_{\rm \alpha}$ cos $\delta$\tablenotemark{b}} & \colhead{$\mu_{\rm \delta}$\tablenotemark{b}} & \colhead{v$_{\rm rad}$\tablenotemark{c}} & \colhead{$\pi$\tablenotemark{d}} & \colhead{$I$\tablenotemark{e}} & \colhead{$J$} & \colhead{X-ray\tablenotemark{f}} & \colhead{$U$} & \colhead{$V$} & \colhead{$W$} & \colhead{X} & \colhead{Y} & \colhead{Z} \\
\colhead{Name} & \colhead{Spt} & \colhead{{hh} {mm} {ss.ss}} & \colhead{{dd} {mm} {ss.s}} & \colhead{(mas yr$^{-1}$)} & \colhead{(mas yr$^{-1}$)} & \colhead{(km s$^{-1}$)} & \colhead{(mas)} & \colhead{(mag)} & \colhead{(mag)} & \colhead{log(f$_{x}$)} & \colhead{(km s$^{-1}$)} & \colhead{(km s$^{-1}$)} & \colhead{(km s$^{-1}$)} & \colhead{(pc)} & \colhead{(pc)} & \colhead{(pc)}
}
\startdata
\cutinhead{$\beta$ Pictoris Moving Group}
 HIP 560  &  F2IV   & $00 $ 06 $ 50.08$ & $-23 $ 06 $ 27.1$ & $  97.81\pm 0.42$ & $ -47.12\pm 0.21$ & $  6.50\pm 3.50$\tablenotemark{n} & $  25.39\pm 0.38$ & $   5.74$ & $  5.45\pm 0.02$ & $-12.36$ & $   -11.0\pm  0.4$ & $   -15.1\pm  0.6$ & $   -10.2\pm  3.4$ & $     4.5$ & $     5.9$ & $   -38.7$ \\
 HIP 10679  &  G2V   & $02 $ 17 $ 24.72$ & $+28 $ 44 $ 30.5$ & $  80.15\pm 4.38$ & $ -78.40\pm 4.91$ & $  4.99\pm 1.26$\tablenotemark{n} & $  36.58\pm 5.83$ & $   7.06$ & $  6.57\pm 0.02$ & $-11.53$ & $    -9.0\pm  1.3$ & $   -10.1\pm  2.2$ & $    -7.2\pm  1.1$ & $   -19.3$ & $    13.6$ & $   -13.9$ \\
 HIP 10680  &  F5V   & $02 $ 17 $ 25.27$ & $+28 $ 44 $ 42.3$ & $  87.60\pm 2.12$ & $ -72.40\pm 2.46$ & $  4.87\pm 1.37$\tablenotemark{n} & $  28.97\pm 2.88$ & $   6.40$ & $  6.05\pm 0.03$ & \ldots  & $   -11.4\pm  1.3$ & $   -13.7\pm  1.8$ & $    -7.2\pm  0.9$ & $   -24.3$ & $    17.1$ & $   -17.5$ \\
 HIP 11152  &  M3Ve   & $02 $ 23 $ 26.63$ & $+22 $ 44 $ 06.9$ & $  92.43\pm 3.05$ & $-113.69\pm 2.36$ & $ 10.40\pm 2.00$\tablenotemark{y} & $  34.86\pm 2.84$ & $   9.44$\tablenotemark{ll} & $  8.18\pm 0.02$ & $-11.41$ & $   -12.2\pm  1.5$ & $   -13.6\pm  1.7$ & $   -13.1\pm  1.3$ & $   -20.1$ & $    12.0$ & $   -16.6$ \\
 HIP 11360\tablenotemark{a} &  F4IV   & $02 $ 26 $ 16.25$ & $+06 $ 17 $ 33.1$ & $  86.31\pm 0.71$ & $ -49.97\pm 0.45$ & $  8.80\pm 3.00$\tablenotemark{s} & $  22.11\pm 0.64$ & $   6.33$ & $  6.03\pm 0.02$ & $-12.18$ & $   -13.3\pm  1.9$ & $   -18.0\pm  0.9$ & $    -5.9\pm  2.3$ & $   -27.8$ & $     9.6$ & $   -34.4$ \\
 HIP 11437 A  &  K8   & $02 $ 27 $ 29.24$ & $+30 $ 58 $ 24.6$ & $  79.78\pm 2.56$ & $ -70.02\pm 1.73$ & $  6.74\pm 0.03$ & $  25.03\pm 2.25$ & $   8.78$ & $  7.87\pm 0.03$ & $-11.37$ & $   -13.4\pm  0.8$ & $   -14.1\pm  1.6$ & $    -8.4\pm  0.6$ & $   -29.4$ & $    19.8$ & $   -18.5$ \\
\enddata
\tablenotetext{a}{{\em bona fide members} with ambiguous membership, see section 9.}
\tablenotetext{b}{Proper motion from \citealp{2007vanleeuwen}, unless stated otherwise.}
\tablenotetext{c}{Radial velocity measurement from \citealp{2012bailey}, unless stated otherwise.}
\tablenotetext{d}{Parallax measurement from \citealp{2007vanleeuwen}, unless stated otherwise.}
\tablenotetext{e}{$I_{c}$ magnitude from \citealp{2011anderson}, unless stated otherwise.}
\tablenotetext{f}{ROSAT X-ray flux calculated from \citet[see Section 5.4;][]{2006riaz} in erg s$^{-1}$ cm$^{-2}$.}
\tablecomments{(g) \citealp{2009teixeira}; (h) \citealp{2008teixeira};
(i) \citealp[][]{2012faherty}; (j) \citealp[PPMXL;][]{2010Roeser}; (k) \citealp{2011riedel};
 (l) \citealp[Riedel (in prep.)][]{}; (m) \citealp[UCAC3;][]{2009zacharias}; (n) \citealp{2007kharchenko}; (o) \citealp{2007bobylev};
 (p) \citealp{2006torres}; (q) \citealp{2001montes}; (r) \citealp{2006bobylev}; (s) \citealp{2006moor};
 (t) \citealp[we add 0.75 mag to the J(2mass) magnitude;][]{2003cutri}; (u) \citealp{2011shkolnik}; (v) \citealp{2010rice}; (w) \citealp{2011kiss};
 (x) \citealp{2012schlieder}; (y) \citealp{2010schlieder}; (z) \citealp{2002mohanty}; (aa) \citealp{1999grenier};
 (bb) \citealp{2000zuckermanwebb}; (cc) \citealp{2011zuckerman}; (dd) \citealp{2007guenther}; (ee) \citealp{2004bzuckerman};
 (ff) \citealp{2006gontcharov}; (gg) \citealp{2011anderson}; (hh) \citealp{2007white}; (ii) \citealp{2007gizis};
 (jj) \citealp[DENIS;][]{1997epchtein}; (kk) \citealp{2010koen}; (ll) \citealp{1987weis}; (mm) \citealp{2010messina}}
\end{deluxetable}
\clearpage
\end{landscape}
\twocolumngrid

\clearpage
\onecolumngrid
\LongTables
\begin{landscape}
\tabletypesize{\scriptsize}
\begin{deluxetable}{lccclccclccclccclccclccclccclccc}
\tablewidth{640pt}
\tablecolumns{32}
\tablecaption{Membership probabilities of all bona fide members\tablenotemark{a} \label{tab:allmemb}}
\tablehead{
\colhead{} & \multicolumn{3}{c}{$\beta$PMG} & \colhead{} & \multicolumn{3}{c}{TWA} & \colhead{} & \multicolumn{3}{c}{THA} & \colhead{} & \multicolumn{3}{c}{COL} &  & \multicolumn{3}{c}{CAR} & \colhead{} & \multicolumn{3}{c}{ARG} & \colhead{} & \multicolumn{3}{c}{ABDMG} & \colhead{} & \multicolumn{3}{c}{Field} \\
\cline{2-4}
\cline{6-8}
\cline{10-12}
\cline{14-16}
\cline{18-20}
\cline{22-24}
\cline{26-28}
\cline{30-32}
\colhead{Name} & \colhead{P} & \colhead{P$_{v}$} & \colhead{P$_{v+\pi}$} & \colhead{} & \colhead{P} & \colhead{P$_{v}$} & \colhead{P$_{v+\pi}$} & \colhead{} & \colhead{P} & \colhead{P$_{v}$} & \colhead{P$_{v+\pi}$} & \colhead{} & \colhead{P} & \colhead{P$_{v}$} & \colhead{P$_{v+\pi}$} & \colhead{} & \colhead{P} & \colhead{P$_{v}$} & \colhead{P$_{v+\pi}$} & \colhead{} & \colhead{P} & \colhead{P$_{v}$} & \colhead{P$_{v+\pi}$} & \colhead{} & \colhead{P} & \colhead{P$_{v}$} & \colhead{P$_{v+\pi}$} & \colhead{} & \colhead{P} & \colhead{P$_{v}$} & \colhead{P$_{v+\pi}$}
}
\startdata
\cutinhead{$\beta$ Pictoris moving group}
 HIP 560 & $ 74$ & $ 98$ & $ 99$ & $ $ & $  0$ & $  0$ & $  0$ & $ $ & $ 15$ & $  0$ & $  0$ & $ $ & $  6$ & $  0$ & $  0$ & $ $ & $  0$ & $  0$ & $  0$ & $ $ & $  0$ & $  0$ & $  0$ & $ $ & $  0$ & $  0$ & $  0$ & $ $ & $  4$ & $  1$ & $  0$ \\
 HIP 10679 & $ 98$ & $ 99$ & $ 99$ & $ $ & $  0$ & $  0$ & $  0$ & $ $ & $  0$ & $  0$ & $  0$ & $ $ & $  0$ & $  0$ & $  0$ & $ $ & $  0$ & $  0$ & $  0$ & $ $ & $  0$ & $  0$ & $  0$ & $ $ & $  0$ & $  0$ & $  0$ & $ $ & $  0$ & $  0$ & $  0$ \\
 HIP 10680 & $ 66$ & $ 99$ & $ 99$ & $ $ & $  0$ & $  0$ & $  0$ & $ $ & $  0$ & $  0$ & $  0$ & $ $ & $ 32$ & $  0$ & $  0$ & $ $ & $  0$ & $  0$ & $  0$ & $ $ & $  0$ & $  0$ & $  0$ & $ $ & $  0$ & $  0$ & $  0$ & $ $ & $  1$ & $  0$ & $  0$ \\
 HIP 11152 & $ 80$ & $ 99$ & $ 99$ & $ $ & $  0$ & $  0$ & $  0$ & $ $ & $  0$ & $  0$ & $  0$ & $ $ & $  0$ & $  0$ & $  0$ & $ $ & $  0$ & $  0$ & $  0$ & $ $ & $  0$ & $  0$ & $  0$ & $ $ & $ 18$ & $  0$ & $  0$ & $ $ & $  0$ & $  0$ & $  0$ \\
 HIP 11360 & $  4$ & $  6$ & $ 12$ & $ $ & $  0$ & $  0$ & $  0$ & $ $ & $  1$ & $  0$ & $  0$ & $ $ & $ 93$ & $ 93$ & $ 87$ & $ $ & $  0$ & $  0$ & $  0$ & $ $ & $  0$ & $  0$ & $  0$ & $ $ & $  0$ & $  0$ & $  0$ & $ $ & $  0$ & $  0$ & $  0$ \\
 HIP 11437 A & $ 65$ & $ 99$ & $ 99$ & $ $ & $  0$ & $  0$ & $  0$ & $ $ & $  0$ & $  0$ & $  0$ & $ $ & $ 33$ & $  0$ & $  0$ & $ $ & $  0$ & $  0$ & $  0$ & $ $ & $  0$ & $  0$ & $  0$ & $ $ & $  0$ & $  0$ & $  0$ & $ $ & $  0$ & $  0$ & $  0$ \\
 HIP 11437 B & $ 77$ & $ 99$ & $ 99$ & $ $ & $  0$ & $  0$ & $  0$ & $ $ & $  0$ & $  0$ & $  0$ & $ $ & $ 22$ & $  0$ & $  0$ & $ $ & $  0$ & $  0$ & $  0$ & $ $ & $  0$ & $  0$ & $  0$ & $ $ & $  0$ & $  0$ & $  0$ & $ $ & $  0$ & $  0$ & $  0$ \\
\enddata
\tablenotetext{a}{Membership probability (P), membership probability including radial velocity information (P$_{v}$), or membership probability including radial velocity and parallax informations (P$_{v+\pi}$).}
\tablenotetext{b}{Membership probability (P or P$_{v}$ or P$_{v+\pi}$) for which the binary hypothesis has a higher probability.}
\end{deluxetable}
\clearpage
\end{landscape}
\twocolumngrid

\clearpage
\onecolumngrid
\LongTables
\begin{landscape}
\tabletypesize{\scriptsize}
\begin{deluxetable}{lrrrrrcrrrrrrrr}
\tablewidth{640pt}
\tablecolumns{15}
\tablecaption{Candidate members of young kinematic groups \label{tab:candprop}}
\tablehead{
\colhead{Name} & \colhead{$I$\tablenotemark{a}} & \colhead{$J$} & \colhead{$\mu_{\rm \alpha}$ cos $\delta$\tablenotemark{b}} & \colhead{$\mu_{\rm \delta}$\tablenotemark{b}} & \colhead{ } & \colhead{X-ray\tablenotemark{c}} & \colhead{H$\alpha$\tablenotemark{d}} & \colhead{Li\tablenotemark{e}} & \colhead{$v_{rad}^{\rm meas}$\tablenotemark{f}} & \colhead{$v_{rad}^{\rm pred}$\tablenotemark{g}} & \colhead{$d_{\pi}$\tablenotemark{h}} & \colhead{$d_{s}$\tablenotemark{g}} & \colhead{P\tablenotemark{i}} & \colhead{Refs.} \\
\colhead{(2MASS)} & \colhead{(mag)} & \colhead{(mag)} & \colhead{(mas yr$^{-1}$)} & \colhead{(mas yr$^{-1}$)} & \colhead{Spt} & \colhead{log(f$_{x}$)} & \colhead{(\AA)} & \colhead{(m \AA)} & \colhead{(km s$^{-1}$)} & \colhead{(km s$^{-1}$)} & \colhead{(pc)} & \colhead{(pc)} & \colhead{(\%)} & \colhead{}
}
\startdata
\cutinhead{$\beta$ Pictoris moving group}
J00172353-6645124 & $ 9.94\pm0.05$ & $ 8.56\pm0.02$ & $ 104.30\pm 1.00$ & $ -13.50\pm 1.00$ &      M2.5V\tablenotemark{}  & $  -11.77$ & $     6.0$ &\ldots  & $  11.4\pm 0.8$\tablenotemark{o} & $  11.0\pm  1.6$ & \ldots  & $  35\pm  3$ & $ 99.9$ &       16  \\[0.5ex]
J01112542+1526214 & $10.98\pm0.00$\tablenotemark{r} & $ 9.08\pm0.03$ & $ 180.00\pm 2.00$\tablenotemark{y} & $-120.00\pm 5.00$\tablenotemark{y} &    M5V+M6V\tablenotemark{}  & $  -12.13$ & $     9.3$ & $     629$\tablenotemark{o} & $   4.0\pm 0.1$ & $   3.1\pm  1.6$ & $  21.8\pm  0.8$\tablenotemark{vv} & $  20\pm  1$ & $ 99.9$\,\tablenotemark{j} &       16  \\[0.5ex]
J01351393-0712517 & $10.53\pm0.08$ & $ 8.96\pm0.02$ & $  96.00\pm 3.00$\tablenotemark{y} & $ -50.00\pm 4.00$\tablenotemark{y} &        M4V\tablenotemark{}  & $  -11.92$ & $    15.3$ &\ldots  & $  11.7\pm 5.3$\tablenotemark{uu} & $   9.4\pm  1.6$ & $  37.9\pm  2.4$\tablenotemark{uu} & $  35\pm  3$ & $ 33.2$\,\tablenotemark{j} &       16  \\[0.5ex]
J01365516-0647379 & $11.22\pm0.06$ & $ 9.71\pm0.02$ & $ 175.10\pm 4.40$\tablenotemark{y} & $ -98.80\pm 4.40$\tablenotemark{y} &        M4V\tablenotemark{}  & $  -12.03$ & $     6.0$ &\ldots  & $  12.2\pm 0.4$\tablenotemark{uu} & $   9.5\pm  1.6$ & $  24.0\pm  0.4$\tablenotemark{uu} & $  21\pm  1$ & $ 27.4$ &       16  \\[0.5ex]
J01535076-1459503 & $ 9.82\pm0.06$ & $ 7.94\pm0.03$ & $ 105.00\pm 1.40$ & $ -45.70\pm 1.40$ &    M3V+M3V\tablenotemark{}  & $  -11.54$ & $     6.6$ &\ldots  & \ldots  & $  12.0\pm  1.6$ & \ldots  & $  28\pm  2$ & $ 99.9$\,\tablenotemark{j} &       16  \\
\enddata
\end{deluxetable}
\clearpage
\end{landscape}
\twocolumngrid

\clearpage
\twocolumngrid
\LongTables
%\begin{landscape}
\tabletypesize{\scriptsize}
\begin{deluxetable}{lcccc}
\tablewidth{640pt}
\tablecolumns{5}
\tablecaption{Ambiguous candidate members of young kinematic groups\tablenotemark{a} \label{tab:candall}}
\tablehead{
\colhead{ } & \colhead{Group} & \colhead{P} & \colhead{d$_{s}$} & \colhead{v$_{\rm rad}^{\rm pred}$} \\
\colhead{Name} & \colhead{     } & \colhead{(\%)} & \colhead{(pc)} & \colhead{(km s$^{-1}$)}
}
\startdata
J00281434-3227556 & $\beta$ PMG  & $ 88.7$\,\tablenotemark{b} & $  33\pm  3$ & $   8.4\pm  1.5$ \\[0.5ex]
\nodata & THA  & $ 10.3$\,\tablenotemark{b} & $  38\pm  2$ & $   0.4\pm  1.3$ \\[0.5ex]
J00551501+3015156 & $\beta$ PMG  & $  9.9$ & $  52\pm  4$ & $  -1.2\pm  1.6$ \\[0.5ex]
\nodata & COL  & $ 80.2$\,\tablenotemark{b} & $  61\pm  2$ & $  -6.1\pm  1.1$ \\[0.5ex]
J01071194-1935359 & $\beta$ PMG  & $  9.1$\,\tablenotemark{b} & $  43\pm  4$ & $   9.3\pm  1.5$ \\[0.5ex]
\nodata & THA  & $  1.6$\,\tablenotemark{b} & $  52\pm  3$ & $   0.6\pm  1.3$ \\[0.5ex]
\nodata & COL  & $ 85.2$\,\tablenotemark{b} & $  68\pm  3$ & $   5.5\pm  0.9$ \\[0.5ex]
J01132817-3821024 & $\beta$ PMG  & $ 22.2$ & $  29\pm  2$ & $  11.9\pm  1.5$ \\
\enddata
\tablenotetext{a}{Membership probability without considering the RV or parallax information (P).}
\tablenotetext{b}{Membership probability (P) for which the binary hypothesis has a higher probability.}
\end{deluxetable}
\clearpage
%\end{landscape}
\twocolumngrid

\clearpage
\onecolumngrid
\LongTables
\begin{landscape}
\tabletypesize{\scriptsize}
\begin{deluxetable}{lccclccclccclccclccclccclccclccc}
\tablewidth{640pt}
\tablecolumns{32}
\tablecaption{Membership probabilities of all candidates\tablenotemark{a} \label{tab:allcand}}
\tablehead{
\colhead{} & \multicolumn{3}{c}{$\beta$PMG} & \colhead{} & \multicolumn{3}{c}{TWA} & \colhead{} & \multicolumn{3}{c}{THA} & \colhead{} & \multicolumn{3}{c}{COL} &  & \multicolumn{3}{c}{CAR} & \colhead{} & \multicolumn{3}{c}{ARG} & \colhead{} & \multicolumn{3}{c}{ABDMG} & \colhead{} & \multicolumn{3}{c}{Field} \\
\cline{2-4}
\cline{6-8}
\cline{10-12}
\cline{14-16}
\cline{18-20}
\cline{22-24}
\cline{26-28}
\cline{30-32}
\colhead{Name} & \colhead{P} & \colhead{P$_{v}$} & \colhead{P$_{v+\pi}$} & \colhead{} & \colhead{P} & \colhead{P$_{v}$} & \colhead{P$_{v+\pi}$} & \colhead{} & \colhead{P} & \colhead{P$_{v}$} & \colhead{P$_{v+\pi}$} & \colhead{} & \colhead{P} & \colhead{P$_{v}$} & \colhead{P$_{v+\pi}$} & \colhead{} & \colhead{P} & \colhead{P$_{v}$} & \colhead{P$_{v+\pi}$} & \colhead{} & \colhead{P} & \colhead{P$_{v}$} & \colhead{P$_{v+\pi}$} & \colhead{} & \colhead{P} & \colhead{P$_{v}$} & \colhead{P$_{v+\pi}$} & \colhead{} & \colhead{P} & \colhead{P$_{v}$} & \colhead{P$_{v+\pi}$}
}
\startdata
J00171443-7032021 & $  0.0$ & \ldots  & \ldots  & $ $ & $  0.0$ & \ldots  & \ldots  & $ $ & $ 99.2$\tablenotemark{b} & \ldots  & \ldots  & $ $ & $  0.0$ & \ldots  & \ldots  & $ $ & $  0.0$ & \ldots  & \ldots  & $ $ & $  0.0$ & \ldots  & \ldots  & $ $ & $  0.5$ & \ldots  & \ldots  & $ $ & $  0.3$ & \ldots  & \ldots  \\[0.5ex]
J00172353-6645124 & $ 99.9$ & $ 99.9$ & \ldots  & $ $ & $  0.0$ & $  0.0$ & \ldots  & $ $ & $  0.0$ & $  0.0$ & \ldots  & $ $ & $  0.0$ & $  0.0$ & \ldots  & $ $ & $  0.0$ & $  0.0$ & \ldots  & $ $ & $  0.4$ & $  0.0$ & \ldots  & $ $ & $  0.1$ & $  0.0$ & \ldots  & $ $ & $  0.0$ & $  0.0$ & \ldots  \\[0.5ex]
J00233468+2014282 & $ 20.9$ & $ 10.3$ & \ldots  & $ $ & $  0.0$ & $  0.0$ & \ldots  & $ $ & $  0.0$ & $  0.0$ & \ldots  & $ $ & $ 18.1$\tablenotemark{b} & $ 35.0$\tablenotemark{b} & \ldots  & $ $ & $  0.0$ & $  0.0$ & \ldots  & $ $ & $  0.0$ & $  0.0$ & \ldots  & $ $ & $  0.0$ & $  0.0$ & \ldots  & $ $ & $ 61.1$ & $ 54.7$ & \ldots  \\[0.5ex]
J00281434-3227556 & $ 88.7$\tablenotemark{b} & \ldots  & \ldots  & $ $ & $  0.0$ & \ldots  & \ldots  & $ $ & $ 10.3$\tablenotemark{b} & \ldots  & \ldots  & $ $ & $  0.9$ & \ldots  & \ldots  & $ $ & $  0.0$ & \ldots  & \ldots  & $ $ & $  0.0$ & \ldots  & \ldots  & $ $ & $  0.0$ & \ldots  & \ldots  & $ $ & $  0.1$ & \ldots  & \ldots  \\[0.5ex]
J00340843+2523498 & $  0.0$ & $  0.0$ & \ldots  & $ $ & $  0.0$ & $  0.0$ & \ldots  & $ $ & $  0.0$ & $  0.0$ & \ldots  & $ $ & $  0.0$ & $  0.0$ & \ldots  & $ $ & $  0.0$ & $  0.0$ & \ldots  & $ $ & $  0.0$ & $  0.0$ & \ldots  & $ $ & $ 98.9$ & $ 97.7$ & \ldots  & $ $ & $  1.1$ & $  2.3$ & \ldots  \\[0.5ex]
J00503319+2449009 & $  0.0$ & $  0.0$ & $  0.0$ & $ $ & $  0.0$ & $  0.0$ & $  0.0$ & $ $ & $  0.0$ & $  0.0$ & $  0.0$ & $ $ & $  0.0$ & $  0.0$ & $  0.0$ & $ $ & $  0.0$ & $  0.0$ & $  0.0$ & $ $ & $ 99.9$\tablenotemark{b} & $ 99.9$\tablenotemark{b} & $  0.0$ & $ $ & $  0.0$ & $  0.0$ & $  0.0$ & $ $ & $  0.3$ & $  0.1$ & $ 99.9$ \\
\enddata
\tablenotetext{a}{Membership probability (P), membership probability including radial velocity information (P$_{v}$), or membership probability including radial velocity and parallax informations (P$_{v+\pi}$).}
\tablenotetext{b}{Membership probability (P or P$_{v}$ or P$_{v+\pi}$) for which the binary hypothesis has a higher probability.}
\end{deluxetable}
\clearpage
\end{landscape}
\twocolumngrid

\clearpage
\twocolumngrid
\LongTables
%\begin{landscape}
\tabletypesize{\scriptsize}
\begin{deluxetable}{lccccc}
\tablewidth{640pt}
\tablecolumns{6}
\tablecaption{Candidate members with parallax measurement \label{tab:plxcand}}
\tablehead{
\colhead{ } & \colhead{d$_{s}$\tablenotemark{c}} & \colhead{d$_{\pi}$\tablenotemark{d}} & \colhead{P$_{\rm v}$\tablenotemark{a}} & \colhead{P$_{\rm v+\pi}$\tablenotemark{a}} & \colhead{ } \\
\colhead{Name} & \colhead{(pc)} & \colhead{(pc)} & \colhead{(\%)} & \colhead{(\%)} & \colhead{Group}
}
\startdata
J00503319+2449009 & $  22.5\pm  1.3$ & $  11.8\pm  0.7$\tablenotemark{f} & $99.99$\tablenotemark{b} & $ 0.00$ &    \\[0.5ex]
J01034210+4051158 & $  33.5\pm  1.6$ & $  29.9\pm  2.0$\tablenotemark{} & $95.64$ & $96.67$ & ABDMG \\[0.5ex]
J01112542+1526214 & $  20.5\pm  1.5$ & $  21.8\pm  0.8$\tablenotemark{e} & $99.99$\tablenotemark{b} & $99.99$\tablenotemark{b} & $\beta$ PMG \\[0.5ex]
J01351393-0712517 & $  35.5\pm  3.1$ & $  37.9\pm  2.4$\tablenotemark{} & $99.99$\tablenotemark{b} & $99.99$\tablenotemark{b} & $\beta$ PMG \\[0.5ex]
J01365516-0647379 & $  21.1\pm  1.7$ & $  24.0\pm  0.4$\tablenotemark{} & $99.99$ & $99.99$ & $\beta$ PMG \\[0.5ex]
J04522441-1649219 & $  16.0\pm  1.2$ & $  16.3\pm  0.4$\tablenotemark{} & $99.99$\tablenotemark{b} & $99.99$\tablenotemark{b} & ABDMG \\[0.5ex]
J05015881+0958587 & $  38.4\pm  3.9$ & $  24.9\pm  1.3$\tablenotemark{e} & $99.99$\tablenotemark{b} & $99.99$\tablenotemark{b} & $\beta$ PMG \\[0.5ex]
J05064946-2135038 & $  21.9\pm  4.4$ & $  19.2\pm  0.5$\tablenotemark{e} & $99.99$\tablenotemark{b} & $99.99$\tablenotemark{b} & $\beta$ PMG \\
\enddata
\tablenotetext{a}{Membership probability including radial velocity information (P$_{v}$) or membership probability including radial velocity and parallax information (P$_{v+\pi}$).}
\tablenotetext{b}{Membership probability (P$_{v}$ or P$_{v+\pi}$) for which the binary hypothesis has a higher probability.}
\tablenotetext{c}{Statistical distance including radial velocity information.}
\tablenotetext{d}{Trigonometric distance from \citet{2012shkolnik}, unless stated otherwise.}
\tablenotetext{e}{Trigonometric distance from Riedel (in prep).}
\tablenotetext{f}{Trigonometric distance from \citet{2007vanleeuwen}.}
\tablenotetext{g}{Trigonometric distance from \citet{2011wahhaj}.}
\end{deluxetable}
\clearpage
%\end{landscape}
\twocolumngrid

\end{document}